\documentstyle[aps,pre,twocolumn,graphicx,psfig]{revtex}

\def\lessgtr{\raise2.5pt\hbox{$<$}\llap{\lower2.5pt\hbox{$>$}}}
\def\gtrless{\raise2.5pt\hbox{$>$}\llap{\lower2.5pt\hbox{$<$}}}

\newcommand{\be}{\begin{equation}}
\newcommand{\ee}{\end{equation}}
\newcommand{\bea}{\begin{eqnarray}}
\newcommand{\eea}{\end{eqnarray}}

\begin{document}

\draft

\title{A mode-coupling theory for the glassy dynamics of 
a diatomic probe molecule immersed in a simple liquid}
\author{S.-H. Chong, W.~G{\"o}tze, and A.P. Singh}
\address{Physik-Department, Technische Universit{\"a}t M{\"u}nchen,
85747 Garching, Germany}
\date{Phys. Rev. E, in print}
\maketitle

\begin{abstract}

Generalizing the mode-coupling theory for ideal
liquid-glass transitions, equations of motion are derived for the
correlation functions describing the glassy dynamics of a diatomic
probe molecule immersed in a simple glass-forming system. The
molecule is described in the interaction-site representation and
the equations are solved for a dumbbell molecule
consisting of two fused hard spheres in a hard-sphere system. The
results for the molecule's arrested position in the glass state
and the reorientational correlators for angular-momentum index
$\ell = 1$ and $\ell = 2$ near the glass transition are compared
with those obtained previously within a
theory based on a tensor-density description of the molecule in
order to demonstrate that the two approaches yield
equivalent results. 
For strongly hindered reorientational motion, the dipole-relaxation
spectra for the $\alpha$-process can be mapped on the dielectric-loss
spectra of glycerol if a rescaling is performed according to 
a suggestion by Dixon {\it et al.} 
[Phys.~Rev.~Lett. {\bf 65}, 1108 (1990)].
It is demonstrated that the glassy dynamics is
independent of the molecule's inertia parameters.

\bigskip

\end{abstract}

\section{INTRODUCTION}

The mode-coupling theory (MCT) for the evolution of structural
relaxation in glass-forming liquids was originally developed
for atomic systems and for mixtures of atoms or ions. Detailed
tests of the theory have been provided by comparisons of the 
predictions for the hard-sphere system (HSS) with 
dynamic-light-scattering data for
hard-sphere colloids, as can be inferred from Ref.~\cite{Megen95}
and the papers quoted there. Quantitative tests have also been
made by comparing molecular-dynamics-simulation data for a
binary mixture with the MCT results for the model
\cite{Nauroth97,Gleim98,Gleim00}. A series of general 
implications of the MCT equations were derived like
scaling laws and relations between anomalous exponents describing
power-law spectra and relaxation-time scales which
establish some universal features of the dynamics~\cite{Goetze92}.
It was conjectured that these 
results also apply to molecular liquids. Indeed, there is a large
body of literature, which is reviewed in Ref.~\cite{Goetze99},
dealing with the analysis of data from experiments or
from molecular-dynamics simulations for complicated systems in
terms of the universal MCT formulas. These studies
suggest that MCT describes some essential features of the glassy
dynamics for molecular liquids. Therefore, it seems desirable
to develop a detailed microscopic theory also for systems of
non-spherical constituents.

A mode-coupling theory for molecular systems has been studied in
Refs.~\cite{Schilling97,Theis98,Fabbian99b,Winkler00,Theis00,Letz00} where
the structure is described by tensor-density
fluctuations. The basic concepts of the MCT for simple systems
like density correlators and relaxation kernels have been
generalized to infinite matrices. The equations for the non-ergodicity
parameters and critical amplitudes were solved. These
quantities generalize the Debye-Waller factors of the arrested
glass structure and characterize its changes with temperature.
Comparison of the theoretical findings with
molecular-dynamics-simulation data for 
water~\cite{Fabbian99b,Theis00}
and for a system of linear molecules \cite{Winkler00} demonstrates
that the theory can cope with microscopic details.
However, the derived equations 
are so involved that further simplifications would be
required before correlators or spectra
could actually be calculated.

The simplest question of glassy dynamics of the rotational
degrees of freedom concerns the motion of a single linear molecule in a
simple liquid. This problem is equivalent to the study of a dilute
solution of linear molecules in an atomic liquid as solvent. For
this system, a MCT has been developed, generalizing the
equation for a tagged particle in a simple liquid to an 
infinite-matrix equation for a tagged molecule \cite{Franosch97c}. 
The equations were solved for a molecule
consisting of two fused hard spheres immersed in a HSS
\cite{Franosch97c,Goetze00c}. The validity of the
universal laws for the reorientational dynamics was
demonstrated. Characteristic differences for the
$\alpha$-process of the relaxation for angular-momentum index
$\ell = 1$ and $\ell = 2$ were identified 
which explain the differences of
spectra measured for $\ell = 1$ by dielectric-loss spectroscopy
and for $\ell = 2$ by depolarized-light-scattering spectroscopy.
The experimentally established large ratio of the
$\alpha$-relaxation times for the $\ell = 2$--reorientational
process and the 
longitudinal elastic modulus was also obtained \cite{Goetze00c}.
These examples show that MCT can provide general insight into the
glassy dynamics of rotational degrees of freedom which goes beyond
the contents of the universal formulas. 

Within the basic version of MCT, 
the tagged-particle-density-fluctuation correlator
for wave number $q$ considered as function of time $t$, 
$\phi_q^s (t)$, or the dynamical structure
factor for frequency $\omega$, $S_{q}(\omega)$, can be written as:
$\phi_q^s (t) = \phi_{q}^{s \, *} (t/t_0)$ and 
$S_{q}(\omega) = S_{q}^{*} (\omega t_0)$. 
Here the functions $\phi_{q}^{s \, *} (\tilde t)$ and 
$S_{q}^{*} (\tilde \omega)$ are completely
determined by the equilibrium structure. 
This holds for times outside the transient regime, $t /
t_0 \gg 1$, or for frequencies below the band of microscopic
excitations, $\omega t_0 \ll 1$. The subtleties of the transient
dynamics like the dependence of oscillation frequencies on mass
ratios enter the long-time dynamics and the low-frequency
spectra via a common time scale $t_0$ only. This means that the
statistical information on the long-time dynamics is 
determined up to a scale $t_0$ by the statistics of the system's
orbits in configuration space rather than 
by the orbits in phase space. The glassy
dynamics as described by functions like $\phi_q^{s} (t)$ or 
$S_{q}(\omega)$ deals with the probabilities of paths through the
high-dimensional potential-energy landscape. The complicated dynamics on
microscopic time scales is irrelevant in the
long-time regime; it merely determines the scale $t_0$ for the
exploration of the configuration space. 
The cited results of the MCT for simple systems and mixtures
\cite{Goetze92,Franosch98,Fuchs99b} are not valid for the
mentioned theories for molecular systems
\cite{Schilling97,Theis98,Fabbian99b,Winkler00,Theis00,Franosch97c,Goetze00c},
which imply isotope effects for the glassy dynamics.
A change of the mass ratio of the molecule's constituents shifts
the center of gravity, and thereby the mode-coupling coefficients
are altered. This leads to shifts of the glass-transition
temperature, the particle's localization lengths, and the like. 
In this respect a system of A-B
molecules would behave qualitatively different than an A-B
mixture. There are no experimental observations
demanding that the long-time dynamics is independent of the
inertia parameters of the molecules. But we
consider the specified isotope effects as artifacts of the
approximations underlying the so far studied extensions of MCT.
This critique and the formidable complexity of the theories based on the
tensor-density descriptions appear as a motivation to search for
an alternative approach towards the glassy dynamics of molecular
systems. An alternative MCT was proposed by 
Kawasaki \cite{Kawasaki97}. But so far, nothing is known about the
solutions of his equations nor on the results concerning
the inertia-parameter issue. In this paper the
suggestion of Chong and Hirata \cite{Chong98b} will be
followed, and the MCT will be based on the interaction-site
representation of the system \cite{Chandler72,Hansen86}.

The description of a molecular liquid by interaction-site
densities is inferior to the one by tensor densities. The
correlators of tensor densities can be used to express
the ones of interaction-site densities but not vice
versa. Interaction-site theories have also difficulties to handle
reorientational correlators. Therefore, it is a major issue of
this paper to show, that the indicated ad hoc objections against a
MCT based on an interaction-site representation do not fully
apply if the theory is restricted to a parameter regime where the
cage effect is the dominant mechanism for the dynamics. To proceed,
the same dumbbell-molecule problem shall be studied, which was
analyzed previously \cite{Franosch97c,Goetze00c}. 

The paper is organized as follows. 
The basic equations for the model are introduced in Sec.~\ref{sec:2}.
Then, the MCT for a diatomic molecule
in a simple liquid is formulated in Sec.~\ref{sec:3}. The major problem is
the derivation of formulas for the mode-coupling coefficients.
This will be done within the Mori-Fujisaka formalism, and the
details are presented in Appendix~\ref{appen:B}. 
In Sec.~\ref{sec:4}, the results of
the theory for the dumbbell in a HSS are discussed.
The findings are summarized in Sec.~\ref{sec:5}.

\section{THE MODEL}
\label{sec:2}

A system of $N$ identical atoms distributed with density $\rho$ at
positions $\vec r_\kappa, \, \kappa = 1, \cdots , N$, is
considered as solvent. The structure can be described by the
density fluctuations for wave vectors $\vec q$: 
$\rho_ {\vec q} = \sum_\kappa \exp (i {\vec q} \cdot {\vec r}_\kappa)$. 
The structure factor 
$S_q = \langle \mid \rho_{\vec q} \mid^2 \rangle / N$
provides the simplest information on the equilibrium distribution
of these particles. Here $\langle \cdots \rangle$ denotes canonical
averaging for temperature $T$. Because of isotropy, $S_q$ only
depends on the wave-number $q = \mid \vec q \mid$. The
Ornstein-Zernike equation, $S_q = 1 / [1 - \rho c_q]$, relates
$S_q$ to the direct correlation function $c_q$. 
The structural dynamics is described in a statistical
manner by the normalized density correlators 
$\phi_q (t) = \langle \rho_{\vec q} (t)^*  \rho_{\vec q} \rangle / N S_q$. 
They are real even functions of time $t$ and exhibit the initial
behavior: $\phi_q (t) = 1 - \frac{1}{2} (\Omega_q t)^2 + O (|t|^3)$.
Here $\Omega_q = q v / \sqrt{S_q}$ is the bare phonon dispersion;
$v = \sqrt{k_B T/m}$ denotes the thermal velocity of the particles
with mass $m$ \cite{Hansen86}.

A rigid molecule of two atoms $A$ and $B$ shall be considered as
solute. Let $\vec r_a, \, a = A$ or $B$, denote the position vectors
of the atoms, so that $L = | \vec r_A - \vec r_B |$ denotes
the distance between the two interaction sites. Vector $\vec e =
(\vec r_A - \vec r_B) / L$ abbreviates the axis of the 
molecule. If $m_a$ denotes the mass of atom $a$, the total mass $M
= m_A + m_B$ and the moment of inertia $I = m_A m_B L^2 / M$
determine the thermal velocities $v_T = \sqrt{k_B T / M}$ and $v_R
= \sqrt{k_B T / I}$ for the molecule's translation and rotation,
respectively. Let us introduce also the center-of-mass position
$\vec r_C = (m_A \vec r_A + m_B \vec r_B) / M$ and the coordinates
$z_{a}$ of the atoms along the molecule axis: $z_A = L (m_B /
M), \, z_B = - L (m_A / M)$. The position of the molecule shall be
characterized by the two interaction-site-density fluctuations
\begin{equation}
\rho^{a}_{\vec q} = \exp ( i {\vec q} \cdot {\vec r_{a}} ),
\quad a = A \mbox{ or } B.
\label{eq:rho-def}
\end{equation}
The two-by-two matrix ${\bf w}_q$ of static fluctuation correlations
$w_q^{ab} = \langle \rho_{\vec q}^{a*} \rho_{\vec q}^b \rangle$ is
given by
\begin{equation}
w^{ab}_{q} = \delta^{ab} + (1 - \delta^{ab}) \, j_{0}(qL),
\label{eq:w-def}
\end{equation}
where here and in the following $j_\ell (x)$ denotes the spherical
Bessel function of index $\ell$. 
The solute-solvent interaction is described 
by the pair-correlation function 
$h_q^a = \langle \rho_{\vec q}^{*} \, \rho_{\vec q}^a \rangle / \rho$, 
which is expressed by a
direct correlation function $c_q^a$ \cite{Chandler72}
\begin{equation}
h^{a}_{q} = S_{q} \sum_{b} w^{ab}_{q} \, c^{b}_{q}.
\label{eq:huv-def}
\end{equation}

The dynamics of the molecule shall be characterized by the 
interaction-site-density correlators 
\begin{equation}
F^{ab}_{q}(t) = \langle \rho^{a}_{\vec q}(t)^{*} \rho^{b}_{\vec q} \rangle.
\label{eq:Fab-def}
\end{equation}
These are real even functions of time obeying $F_q^{ab} (t)
= F_q^{ba} (t)$. They shall be combined to a two-by-two-matrix
correlator ${\bf F}_q (t)$. Its short-time expansion can be noted
as
\begin{equation}
{\bf F}_{q}(t) = {\bf w}_{q} - 
{\textstyle \frac{1}{2}} \, q^{2} \, {\bf J}_{q} \, t^{2} +
{\bf O}(|t|^{3}).
\label{eq:Fab-short-time}
\end{equation}
The continuity equation reads 
$\dot \rho_{\vec q}^a = i {\vec q} \cdot {\vec j}_{\vec q}^a$, 
where the current fluctuation is 
$\vec j_{\vec q}^a = \vec v^a \rho_{\vec q}^a$ with $\vec v^a$ denoting the
velocity of atom $a$. Therefore, one gets 
$J_q^{ab} = 
\langle 
({\vec q} \cdot {\vec j}_{\vec q}^a)^* ({\vec q} \cdot {\vec j}_{\vec q}^b) 
\rangle / q^2$. 
The result splits in a translational and rotational part, 
${\bf J}_q = {\bf J}^{\rm T}_{q} + {\bf J}^{\rm R}_{q}$, 
where~\cite{Chong98}:
\begin{mathletters}
\label{eq:J-TR}
\bea
J^{{\rm T} \, ab}_{q} &=& v_{\rm T}^{2} \, w^{ab}_{q},
\label{eq:J-TR-a}
\\
J^{{\rm R} \, ab}_{q} &=& v_{\rm R}^{2} \, 
( {\textstyle \frac{2}{3}} z_{a} z_{b} ) \,
[ \delta^{ab} + 
\nonumber \\
& & \qquad \qquad \qquad
(1-\delta^{ab}) \, (j_{0}(qL) + j_{2}(qL)) ].
\label{eq:J-TR-b}
\eea

Let us note the small-$q$ expansion of the density correlators in
the form
\end{mathletters}
\begin{equation}
F^{ab}_{q}(t) = 1 - {\textstyle \frac{1}{6}} \, q^{2} \, C^{ab}(t) + O(q^{4}). 
\label{eq:Fab-small-q}
\end{equation}
The diagonal elements of the symmetric matrix ${\bf C}(t)$ are the
mean-squared displacements
\begin{mathletters}
\label{eq:Cab-mat}
\begin{equation}
\delta r_{a}^{2}(t) = 
\langle ( {\vec r}_{a}(t) - {\vec r}_{a}(0) )^{2} \rangle =
C^{aa}(t),
\label{eq:Cab-mat-a}
\end{equation}
while the off-diagonal elements can be related to the
dipole correlator
\bea
C_{1}(t) &=& \langle {\vec e}(t) \cdot {\vec e} \, \rangle
\nonumber \\
&=&
[ C^{AB}(t) - {\textstyle \frac{1}{2}} ( C^{AA}(t) + C^{BB}(t) ) ] \, / \, L^{2}.
\label{eq:Cab-mat-b}
\eea
The mean-squared displacement of the center of mass can be
expressed as 
\bea
\delta r_{C}^{2}(t) &=& 
\langle ( {\vec r}_{C}(t) - {\vec r}_{C}(0) )^{2} \rangle
\nonumber \\
&=&
[ m_{A} \delta r_{A}^{2}(t) + m_{B} \delta r_{B}^{2}(t)] / M 
\nonumber \\
& &
\qquad \qquad \qquad 
+ \, 
(2 I / M) \, [ C_{1}(t) - 1].
\label{eq:Cab-mat-c}
\eea
Expanding Eq.~(\ref{eq:Fab-short-time}) in $q$ yields the initial decay
\end{mathletters}
\begin{equation}
{\bf C}(t) = {\bf C}_{0} + 3 \, {\bf J}_{0} \, t^{2} + {\bf O}(|t|^{3}).
\label{eq:Cab-short-time}
\end{equation}
Here the initial value ${\bf C}_{0}$ is due to the expansion of 
Eq.~(\ref{eq:w-def}), 
while the prefactor of the $t^2$-term is due to the
zero-wave-number limit of Eqs.~(\ref{eq:J-TR}):
\begin{equation}
C^{ab}_{0} = L^{2} \, (1 - \delta^{ab}), \quad
J^{ab}_{0} = v_{\rm T}^{2} + {\textstyle \frac{2}{3}} \, 
v_{\rm R}^{2} \, z_{a} \, z_{b}.
\label{eq:Cab0-Jab0}
\end{equation}

For symmetric molecules one gets $m_{A} = m_{B} = M/2$,
$I = ML^{2}/4$, and $z_{A} = - z_{B} = L/2$.
In this case, there are only two independent density correlators,
since $F^{AA}_{q}(t) = F^{BB}_{q}(t)$.
It is convenient to perform an orthogonal transformation to fluctuations
of total number-densities $\rho_{N}({\vec q} \,)$ and 
``charge'' densities $\rho_{Z}({\vec q} \,)$:
\begin{mathletters}
\label{eq:rho-NZ}
\begin{equation}
\rho_{x}({\vec q} \,) = 
( \rho^{A}_{\vec q} \pm \rho^{B}_{\vec q} ) \, / \, \sqrt{2}, \quad
x = N \mbox{ or } Z.
\label{eq:rho-NZ-a}
\end{equation}
The transformation matrix ${\bf P} = {\bf P}^{-1}$ reads
\begin{equation}
{\bf P} = \frac{1}{\sqrt{2}}
\left(
\begin{array}{rr}
1 &  1 \\
1 & -1 
\end{array}
\right).
\label{eq:rho-NZ-b}
\end{equation}
It diagonalizes the matrices ${\bf w}_{q}$ from Eq.~(\ref{eq:w-def})
and ${\bf J}_{q}$ from Eqs.~(\ref{eq:J-TR}):
\bea
( {\bf P} \, {\bf w}_{q} \, {\bf P} )^{xy} &=& 
\delta^{xy} \, w_{x}(q), \quad
w_{x}(q) = 1 \pm j_{0}(qL),
\label{eq:rho-NZ-c}
\\
( {\bf P} \, {\bf J}^{\rm T}_{q} \, {\bf P} )^{xy} &=& 
\delta^{xy} \, v_{\rm T}^{2} \, w_{x}(q),
\label{eq:rho-NZ-d}
\\
( {\bf P} \, {\bf J}^{\rm R}_{q} \, {\bf P} )^{xy} &=& 
\delta^{xy} \, {\textstyle \frac{1}{6}} \, v_{\rm R}^{2} \, L^{2} \,
[1 \mp (j_{0}(qL) + j_{2}(qL))],
\label{eq:rho-NZ-e}
\eea
where $x,y = N$ or $Z$.
\end{mathletters}
Also the matrix of density correlators is diagonalized.
Introducing the normalized correlators
$\phi_{q}^{x}(t)$, one gets
\begin{mathletters}
\label{eq:phi-NZ}
\bea
& &
\phi^{x}_{q}(t) =
\langle \rho_{x}({\vec q},t)^{*} \rho_{x}({\vec q}) \rangle \, / \, w_{x}(q), 
\\
& &
( {\bf P} \, {\bf F}_{q}(t) \, {\bf P} )^{xy} =
\delta^{xy} \, \phi^{x}_{q}(t) \, w_{x}(q).
\eea
The mean-squared displacements are equal and shall be denoted by
$\delta r^{2}(t) = \delta r_{A}^{2}(t) = \delta r_{B}^{2}(t)$,
so that Eq.~(\ref{eq:Cab-mat-c}) reads
$\delta r^{2}(t) = \delta r_{C}^{2}(t) + 
(1/2) L^{2} [1 - C_{1}(t)]$.
\end{mathletters} 
The matrix ${\bf C}(t)$ is diagonalized 
\begin{mathletters}
\label{eq:Cmat-symm}
\bea
( {\bf P} \, {\bf C}(t) \, {\bf P} )^{NN} &=&
2 \delta r_{C}^{2}(t) + L^{2}, 
\\
( {\bf P} \, {\bf C}(t) \, {\bf P} )^{ZZ} &=&
- L^{2} C_{1}(t). 
\eea

In Appendix A it is shown how the correlation functions in the
interaction-site representation can be expressed in terms of the
ones in the tensor-density representation.
\end{mathletters}

\section{APPROXIMATIONS}
\label{sec:3}

\subsection{The solvent-density correlator}
\label{subsec:MCT-v}

The density correlator of the solvent is needed to formulate the
equations for the probe molecule. This quantity is discussed
comprehensively in the preceding literature on the MCT for simple
systems~\cite{Franosch97}. 
Let us note here only those equations which have to
be solved in order to obtain the input information for
the calculations of the present paper. First, there is the exact
Zwanzig-Mori equation of motion \cite{Hansen86} relating the
correlator for density fluctuations $\phi_q (t)$ to the correlator
$m_q (t)$ for the force fluctuations:
\begin{equation}
\partial_{t}^{2} \phi_{q}(t) + \Omega_{q}^{2} \, \phi_{q}(t) +
\Omega_{q}^{2} \int_{0}^{t} dt' \, m_{q}(t-t') \, \partial_{t'} \phi_{q}(t') = 0.
\label{eq:GLE-v}
\end{equation}
Second, there is the approximate expression for kernel $m_q (t)$
as mode-coupling functional
\begin{mathletters}
\label{eq:MCT-v}
\begin{equation}
m_{q}(t) = {\cal F}_{q}[\phi(t)].
\label{eq:MCT-v-a}
\end{equation}
The functional ${\cal F}_q$ is rederived as Eq.~(\ref{eq:BXX-1}) in 
Appendix~\ref{appen:B}. 
The wave numbers are discretized to $M$ values with spacing
$h$: $q/h = 1/2, \, 3/2, \cdots, M-1/2$. Then $\phi (t)$
and similar quantities are to be viewed as vectors of $M$
components $\phi_q (t)$, $q=1,\cdots, M$, and the functional is 
\begin{equation}
{\cal F}_{q}[\tilde{f}] = \sum_{kp} V_{q,kp} \, \tilde{f}_{k} \, \tilde{f}_{p}.
\label{eq:MCT-v-b}
\end{equation}
Third, Eqs.~(\ref{eq:GLE-v}) and (\ref{eq:MCT-v-a}) imply the equation for
the long-time limit ${f_q = \phi_q (t \to \infty)}$:
\end{mathletters}
\begin{equation}
f_{q} = {\cal F}_{q}[f] \, / \, \{ 1 + {\cal F}_{q}[f] \}.
\label{eq:DW-v}
\end{equation}
For the liquid state, there is only the trivial solution $f_q =
0$. The glass is characterized by a non-ergodicity
parameter $0 < f_q < 1$, which has the meaning of the Debye-Waller factor of the
arrested structure. At the liquid-glass transition, the long-time
limit of the correlator changes discontinuously from zero to the
critical value $f_q^c > 0$.

\subsection{The solute-interaction-site-density correlators}
\label{subsec:MCT-u}

For matrices of correlation functions as defined in 
Eq.~(\ref{eq:Fab-def}),
the Zwanzig-Mori formalism also leads to an exact equation of motion
\cite{Hansen86}:
\begin{mathletters}
\label{eq:GLE-u}
\begin{equation}
\partial_{t}^{2} {\bf F}_{q}(t) + {\bf \Omega}_{q}^{2} \, {\bf F}_{q}(t) +
{\bf \Omega}_{q}^{2} 
\int_{0}^{t} dt' \, {\bf m}_{q}(t-t') \, \partial_{t'} {\bf F}_{q}(t') = {\bf 0}.
\label{eq:GLE-u-a}
\end{equation}
From the short-time expansion together with Eq.~(\ref{eq:Fab-short-time}), 
one gets
\begin{equation}
{\bf \Omega}_{q}^{2} = q^{2} \, {\bf J}_{q} \, {\bf w}^{-1}_{q}.
\label{eq:GLE-u-b}
\end{equation}
The r.h.s. of this equation is a product of two symmetric positive
definite matrices. 
\end{mathletters}
Hence it can be written as square of a matrix
${\bf \Omega}_q$.
Splitting off this matrix in front of the
convolution integral is done for later convenience.

The difficult problem is the derivation of an approximation for
the matrix ${\bf m}_{q}(t)$ of fluctuating-force correlations, so that the
cage effect is treated reasonably. 
The result, Eq.~(\ref{eq:BXX-2}) from 
Appendix~\ref{appen:B}, can be formulated as mode-coupling
functional 
$\mbox{\boldmath ${\cal F}$}_{q}$:
\begin{mathletters}
\label{eq:MCT-u}
\begin{equation}
m^{ab}_{q}(t) = {\cal F}^{ab}_{q}[{\bf F}(t), \phi(t)]. 
\label{eq:MCT-u-a}
\end{equation}
After the discretization of the wave numbers as explained above,
$\mbox{\boldmath ${\cal F}$}_{q}$ reads
\begin{equation}
{\cal F}^{ab}_{q}[\tilde{\textit{\textbf f}}, \tilde{f}] = q^{-2} 
\sum_{c} w^{ac}_{q} \, \sum_{kp} V^{cb}_{q,kp} \, 
\tilde{f}^{cb}_{k} \, \tilde{f}_{p}. 
\label{eq:MCT-u-b}
\end{equation}
The preceding equations are matrix generalizations of the MCT
equations for the tagged-particle-density correlator $\phi_q^s
(t)$ in a simple liquid \cite{Fuchs98}. 
\end{mathletters}

The equation for the non-ergodicity parameters of the molecule,
$F_q^{ab \, \infty} = F_q^{ab} (t \to \infty)$, can be obtained from
Eqs.~(\ref{eq:GLE-u-a}) and (\ref{eq:MCT-u-a}). 
It is a matrix generalization of Eq.~(\ref{eq:DW-v}):
\begin{equation}
{\bf F}^{\infty}_{q} = 
\mbox{\boldmath ${\cal F}$}_{q}[{\bf F}^{\infty},f] \,
\{ {\bf 1} + \mbox{\boldmath ${\cal F}$}_{q}[{\bf F}^{\infty},f] \}^{-1} \, 
{\bf w}_{q}.
\label{eq:DW-u}
\end{equation}
If the solvent is a liquid, i.e., if $f_q = 0$, one gets
${\bf F}_q^\infty = {\bf 0}$. If the solvent is a glass, 
the long-time limits $F_q^{ab \, \infty}$ can be non trivial. In
this case, the solvent properties enter via the Debye-Waller
factors $f_q$, which renormalize the coupling coefficients 
$V_{q,kp}^{cb}$ in Eq.~(\ref{eq:MCT-u-b}).

Let us specialize to symmetric molecules. 
Multiplying Eqs.~(\ref{eq:GLE-u}) to (\ref{eq:DW-u}) from left and right with 
$\bf P$ given by Eq.~(\ref{eq:rho-NZ-b}) and inserting ${\bf 1} = {\bf P P}$ 
between every pair of matrices, all equations are
transformed to diagonal ones. Thus, there are two equations of
motion
\bea
& &
\partial_{t}^{2} \phi^{x}_{q}(t) + \Omega^{x \, 2}_{q} \, \phi^{x}_{q}(t) 
\nonumber \\
& & 
\qquad \qquad \qquad
+ \,
\Omega^{x \, 2}_{q} 
\int_{0}^{t} dt' \, m^{x}_{q}(t-t') \, \partial_{t'} \phi^{x}_{q}(t') = 0, 
\label{eq:GLE-NZ}
\eea
for $x=N$ or $Z$. 
The two characteristic frequencies $\Omega_q^x$, which specify the
initial decay of the correlators by $\phi_q^x (t) = 1 -
\frac{1}{2} (\Omega_q^x t)^2 + O (|t|^3)$, read:
\begin{mathletters}
\label{eq:Omega-NZ}
\bea
\Omega^{N \, 2}_{q} &=& (v_{\rm T} q)^{2} + 
{\textstyle \frac{1}{6}} 
(v_{\rm R} L q)^{2} [1 - j_{0}(qL) - j_{2}(qL)] 
\nonumber \\
& &
\qquad \qquad \qquad \qquad \qquad
/ \, 
[1 + j_{0}(qL)],
\label{eq:Omega-NZ-a}
\\
\Omega^{Z \, 2}_{q} &=& (v_{\rm T} q)^{2} +
{\textstyle \frac{1}{6}} 
v_{\rm R}^{2} [1 + j_{0}(qL) + j_{2}(qL)] (qL)^{2} 
\nonumber \\
& &
\qquad \qquad \qquad \qquad \qquad
/ \, 
[1 - j_{0}(qL)].
\label{eq:Omega-NZ-b}
\eea
The relaxation kernels can be written as $m_q^x (t) = {\cal F}_q^x
[\phi^x (t), \phi (t)]$, where Eq.~(\ref{eq:BXX-2}) gives 
\end{mathletters}
\begin{mathletters}
\label{eq:MCT-NZ}
\bea
{\cal F}^{x}_{q}[\tilde{f}^{x}, \tilde{f}] &=& [w_{x}(q)/q^{2}] \,
\int \frac{d{\vec k}}{2 (2\pi)^{3}} \,
({\vec q} \cdot {\vec p} / q)^{2} 
\nonumber \\
& &
\qquad \qquad
\times \,
w_{x}(k) \, \rho S_{p} \, c^{N}(p)^{2} \,
\tilde{f}^{x}_{k} \, {\tilde f}_{p}. 
\label{eq:MCT-NZ-a}
\eea
Here ${\vec p} = {\vec q} - {\vec k}$, and 
$c^N (p) = \sqrt{2} c^A_p = \sqrt{2} c^B_p$.
The above specified discretization of the wave numbers yields
${\cal F}_q^x$ as polynomial
\begin{equation}
{\cal F}^{x}_{q}[\tilde{f}^{x}, \tilde{f}] = 
[w_{x}(q)/q^{2}] \,
\sum_{kp} V^{x}_{q, kp} \, \tilde{f}^{x}_{k} \, \tilde{f}_{p}.
\label{eq:MCT-NZ-b}
\end{equation}
One gets for the non-ergodicity parameters $f_q^x = \phi_q^x
(t \to \infty) = (F_q^{AA \, \infty} \pm F_q^{AB \, \infty}) / w_x (q)$
\end{mathletters}
\begin{equation}
f^{x}_{q} = {\cal F}^{x}_{q}[f^x, f] \, / \, 
\{ 1 + {\cal F}^{x}_{q}[f^x, f] \}.
\label{eq:DW-NZ}
\end{equation}

There is no coupling between the fluctuations of the total density
and the ones of the ``charge'' density. The mathematical structure
of the two sets of equations for $x = N$ and $x = Z$, respectively,
is the same as the one studied previously for the density correlator
$\phi_q^s (t)$ of a tagged particle in a simple liquid
\cite{Fuchs98}. For the density dynamics one also finds the
small-$q$ asymptote for the frequency
$\Omega_q^{N \, 2} = (v_{\rm T} q)^2 + O (q^4)$, 
reflecting free translation of the probe molecule.
There also is the $q^{-2}$-divergency of the mode-coupling
coefficients in ${\cal F}_q^N$, which implies the approach towards
unity of the Lamb-M\"ossbauer factor for vanishing wave number:
$f_{q \to 0}^N = 1$. For the ``charge'' dynamics, one gets 
a non-zero small-$q$ limit for the frequency 
characterizing free rotation 
$\Omega_{q \to 0}^{Z \, 2} = 2 v_{\rm R}^2 + O(q^2)$. 
The mode-coupling coefficients do not diverge for $q \to 0$, 
since $6 w_Z (q) / (L q )^2 \to 1$. 
Therefore, the non-ergodicity parameter for the variable 
$\rho_Z ({\vec q},t)$ approaches a limit smaller than unity:
$f^{Z}_{q \to 0} < 1$.

\subsection{The dipole correlator and the mean-squared displacements}
\label{subsec:MCT-C1-MSD}

According to Eqs.~(\ref{eq:Cab-mat}), the knowledge of the dipole correlator 
$C_1(t)$ and two of the mean-squared displacements $\delta r_a^2 (t)$ for $a
= A, B$, or $C$ is equivalent to the knowledge of the three
independent elements of the symmetric matrix ${\bf C} (t)$. Using
Eq.~(\ref{eq:Fab-small-q}) and expanding 
Eq.~(\ref{eq:GLE-u-a}) for small wave numbers one gets
\begin{equation}
\partial_{t}^{2} {\bf C}(t) - {\bf D} + {\bf \Omega}^{2}_{0} {\bf C}(t) +
{\bf J}_{0} 
\int_{0}^{t} dt' {\bf m}(t-t') \partial_{t'} {\bf C}(t') = {\bf 0}.
\label{eq:GLE-Cmat}
\end{equation}
This exact equation of motion for ${\bf C} (t)$ has to be solved
with the initial condition from Eq.~(\ref{eq:Cab-short-time}). 
The frequency matrix is obtained as zero-wave-number limit from 
Eq.~(\ref{eq:GLE-u-b}):
\begin{equation}
{\bf \Omega}^{2}_{0} = (2 v_{\rm R}^{2} / L) \, 
\left(
\begin{array}{rr}
z_{A} & - z_{A} \\
z_{B} & - z_{B} 
\end{array}
\right).
\label{eq:Omega-Cmat}
\end{equation}
Equation~(\ref{eq:GLE-Cmat}) implies 
$\ddot{\bf C}(0) - {\bf D} + {\bf \Omega}_0^2 {\bf C} (0) = {\bf 0}$. 
Thus, one gets from Eq.~(\ref{eq:Cab-short-time}): 
${\bf D} = 6 {\bf J}_0 + {\bf \Omega}_0^2 {\bf C}_0$, i.e., 
$D^{ab} = 6 \, v_{\rm T}^{2} + 2 \, v_{\rm R}^{2} \, (z_{A} + z_{B}) \, z_{a}$.

The MCT approximation for the kernel ${\bf m}(t)$ is obtained by combining
Eqs.~(\ref{eq:GLE-u-b}) and (\ref{eq:MCT-u-b}) and 
taking the zero-wave-vector limit. 
With Eq.~(\ref{eq:BXX-2}), one finds
\begin{mathletters}
\label{eq:MCT-Cmat}
\bea
{\bf m}(t) &=& \mbox{\boldmath ${\cal F}$}[{\bf F}(t), \phi(t)],
\label{eq:MCT-Cmat-a}
\\
{\cal F}^{ab}[\tilde{\textit{\textbf f}}, \tilde{f}] &=& \frac{1}{6 \pi^{2}}
\int_{0}^{\infty} dk \, k^{4} \, \rho S_{k} \, c^{a}_{k} \, c^{b}_k \, 
\tilde{f}^{ab}_{k} \, \tilde{f}_{k}.
\label{eq:MCT-Cmat-b}
\eea

Again, the theory simplifies considerably for symmetric molecules.
In this case, one can transform Eq.~(\ref{eq:GLE-Cmat}) as explained in
connection with the derivation of Eq.~(\ref{eq:GLE-NZ}). 
\end{mathletters}
Using Eqs.~(\ref{eq:Cmat-symm}) one
gets the exact equation of motion for the mean-squared
displacement 
\begin{equation}
\partial_{t}^{2} \delta r^{2}_{C}(t) - 6 v_{\rm T}^{2} + v_{\rm T}^{2}
\int_{0}^{t} dt' \, m_{N}(t-t') \, 
\partial_{t'} \delta r^{2}_{C}(t') = 0,
\label{eq:GLE-MSD}
\end{equation}
to be solved with the initial behavior 
$\delta r_{C}^{2} (t) = 3 (v_{\rm T} t)^2 + O (|t|^3)$. 
Similarly, one obtains for the dipole correlator
\bea
& &
\partial_{t}^{2} C_{1}(t) + 2 v_{\rm R}^{2} \, C_{1}(t) 
\nonumber \\
& & 
\qquad \qquad
+ \, 
2 v_{\rm R}^{2}
\int_{0}^{t} dt' \, m_{Z}(t-t') \, \partial_{t'} C_{1}(t') = 0,
\label{eq:GLE-C1}
\eea
to be solved with the initial decay 
$C_1 (t) = 1 - (v_{\rm R} t)^2 + O(|t|^3)$. 
The MCT approximation for the kernels is obtained from 
Eq.~(\ref{eq:MCT-Cmat-b}):
\begin{mathletters}
\label{eq:MCT-MSD-C1}
\bea
m_{x}(t) &=& {\cal F}_{x}[\phi^{x}(t),\phi(t)], \quad
x = N \mbox{ or } Z,
\label{eq:MCT-MSD-C1-a}
\\
{\cal F}_{x}[\tilde{f}^{x}, \tilde{f}] &=& \alpha_{x} 
\int_{0}^{\infty} dk \,
k^{4} \, \rho S_{k} \, c^{N}(k)^{2} \, w_{x}(k) \, \tilde{f}^{x}_{k} \, \tilde{f}_{k},
\label{eq:MCT-MSD-C1-b}
\eea
where $\alpha_N = 1 / (6 \pi^2)$ and $\alpha_Z = L^2 / (72\pi^2)$. 
\end{mathletters}
Equations~(\ref{eq:GLE-C1}) and (\ref{eq:MCT-MSD-C1}) for the dipole correlator
have the standard form of the MCT equation. If $\phi_q^Z (t)$
approaches zero for large times, the same approach towards
equilibrium is exhibited by $C_1 (t)$. If the solvent is a glass,
$f_q > 0$, and if the ``charge''-density fluctuations
$\phi_q^Z (t)$ exhibit non-ergodic behavior, $f_q^Z > 0$, also the
$\ell = 1$--reorientational correlator exhibits non-ergodic
dynamics:
\bea
C_{1}(t \to \infty) = f_{1} = 
{\cal F}_{Z}[f^{Z}, f] \, / \, \{ 1 + {\cal F}_{Z}[f^{Z}, f] \}.
\label{eq:f1-symm}
\eea
Parameter $f_{1}$ is the long-wave-length limit of $f_{q}^{Z}$
discussed in Eq.~(\ref{eq:DW-NZ}):
$f^{Z}_{q \to 0} = f_{1}$.

\subsection{The quadrupole correlator}
\label{subsec:MCT-C2}

The quadrupole correlator 
$C_2 (t) = \langle 3 [{\vec e}(t) \cdot {\vec e}\,]^{2} - 1 \rangle / 2$ 
cannot be extracted from the correlators $F_q^{a b} (t)$ with $a, b
= A$ or $B$. But let us consider a linear symmetric triatomic
molecule. 
The third atom, labeled $C$, has its position in the
center $\vec r_C$. The preceding theory can be extended by
adding as third variable the fluctuations $\rho_{\vec q}^C = \exp
(i \vec q \cdot \vec r_C)$. The basic quantities are now the 
elements of the 3-by-3 matrix correlator, defined as in 
Eq.~(\ref{eq:Fab-def}) with $a,b = A, B$ or $C$. 
The correlator formed with
$\rho_{Q}({\vec q}) = 
\rho_{\vec q}^A + \rho_{\vec q}^B - 2 \rho_{\vec q}^C$ 
is a linear combination of the nine functions $F_q^{ab} (t)$. 
An expansion for small $q$ yields
\begin{equation}
\langle \rho_{Q}({\vec q},t)^{*} \rho_{Q}({\vec q}) \rangle =
{\textstyle \frac{1}{180}} \, (qL)^{4} \, 
[C_{2}(t) + {\textstyle \frac{5}{4}}] + O(q^{6}).
\label{eq:C2-ABC}
\end{equation}
In this case, $C_2 (t)$ can be obtained in a similar manner as
discussed above for $C_1 (t)$. A diatomic molecule can be
considered as a special mathematical limit of a triatomic one.
Hence, there is no problem in principle to obtain $C_2 (t)$
within a theory based on an interaction-site description.
Motivated by  this observation, an auxiliary site $C$ shall be
introduced \cite{Hoye77,Sullivan81} 
and $\rho_{\vec q}^C$ will be used as third
basic variable. 
However, a complete theory with 3-by-3 matrices shall not
be developed. Rather some additional approximations will be
introduced so that $C_2 (t)$ is obtained
as corollary of the above formulated closed theory.

The quadrupole correlator can be written as small-$q$ limit of a
correlation function formed with tensor-density fluctuations
defined in Eq.~(\ref{eq:rho-tensor-def}) for 
${\vec q}_{0}=(0,0,q)$:
$C_2 (t) = \lim_{q \to 0} 
\langle \rho_2^0 ({\vec q}_0, t)^* \rho_2^0 ({\vec q}_0) \rangle$.
Therefore, an exact Zwanzig-Mori equation can be derived as usual:
\bea
& &
\partial_{t}^{2} C_{2}(t) + 6 v_{\rm R}^{2} \, C_{2}(t) 
\nonumber \\
& & 
\qquad \qquad
+ \, 
6 v_{\rm R}^{2}
\int_{0}^{t} dt' \, m_{2}^{\rm R}(t-t') \, \partial_{t'} C_{2}(t') = 0.
\label{eq:GLE-C2}
\eea
The relaxation kernel $m_2^{\rm R} (t)$ is a correlator for fluctuating
forces $F_{{\rm R} 2} (q 0, t)$ referring to angular-momentum index
$\ell = 2$ and helicity $m = 0$:
\begin{equation}
m_{2}^{\rm R}(t) = \lim_{q \to 0} 
\langle F_{{\rm R} 2}(q0,t)^{*} F_{{\rm R} 2}(q0) \rangle.
\label{eq:m2-def}
\end{equation}
The time evolution of the fluctuating force is generated by the
reduced Liouvillian ${\cal L}^\prime = {\cal Q L Q}$, where $\cal
Q$ projects perpendicular to $\rho_2^0 (\vec q_0)$ and ${\cal L}
\rho_2^0 (\vec q_0)$, and the Liouvillian ${\cal L}$ is
defined by $i {\cal L} A (t) = \partial_t A (t)$.
The involved notation has been chosen in order to get the formulas
in agreement with the ones of the more general theory in 
Ref.~\cite{Franosch97c}. The procedure used for the theory of simple
liquids \cite{Goetze91b} shall be applied to derive an approximation for
the kernel. First, the forces will be approximated by the
projection onto the space of the simplest modes contributing 
$F_{{\rm R} 2} \to {\cal P}' F_{{\rm R} 2}$. 
Here ${\cal P}'$ projects onto the space
spanned by the pair modes
\begin{mathletters}
\label{eq:m2-MCT}
\begin{equation}
A^{a}({\vec k}, {\vec p}) = \rho^{a}_{\vec k} \, \rho_{\vec p} \, / 
\sqrt{N S_{p}}, \quad
a = A, B \mbox{ or } C.
\label{eq:m2-MCT-a}
\end{equation}
The essential step is the second one, where correlations of the
pairs are replaced by products of correlations: 
$\langle A^{a} ({\vec k}, {\vec p}, t)^* A^{a'} ({\vec k}', {\vec p} \, ') \rangle \to 
\delta_{{\vec k} {\vec k}'} \delta_{{\vec p} {\vec p} \, '}
\langle \rho_{\vec k}^a (t)^{*} \rho_{\vec k}^{a'} \rangle \, 
\langle \rho_{\vec p}(t)^{*}    \rho_{\vec p} \rangle / NS_{p}$. 
This approximation is
done in particular for $t = 0$, thereby getting for the
normalization matrix for the pair modes: 
$\langle A^{a} ({\vec k}, {\vec p})^* A^{b} ({\vec k}', {\vec p} \, ') \rangle \to 
\delta_{{\vec k} {\vec k}'} \, \delta_{{\vec p} {\vec p} \, '} \, 
w_k^{ab}$. 
Here $w^{ab} (k) = j_0 (k (z_a - z_b))$ 
generalizes Eq.~(\ref{eq:w-def}) to a 3-by-3 matrix.
As a result, one gets
\bea
m^{\rm R}_{2}(t) &=& \lim_{q \to 0} \sum_{{\vec k} {\vec p}}
\sum_{abcd} \langle F_{{\rm R} 2}(q0)^{*} A^{a}({\vec k},{\vec p} \, ) \rangle \,
({\bf w}_{k}^{-1})^{ab} 
\nonumber \\
&\times&
F_{k}^{bc}(t) \, \phi_{p}(t) \,
({\bf w}_{k}^{-1})^{cd} \, 
\langle A^{d}({\vec k},{\vec p} \, )^{*} F_{{\rm R} 2}(q0) \rangle.
\label{eq:m2-MCT-b}
\eea
The $q \to 0$ limit is carried out easily, reducing the sum over
$\vec k$ and $\vec p$ to the one over $\vec k$ with ${\vec k} = - {\vec p}$.
\end{mathletters}
One obtains the kernel as mode-coupling functional
\begin{equation}
m^{\rm R}_{2}(t) = \int_{0}^{\infty} dk 
\sum_{ab \in \{A, B, C\}} 
V_{2}^{ab}(k) \, F^{ab}_{k}(t) \, \phi_{k}(t). 
\label{eq:m2-final}
\end{equation}
Let us restrict the discussion to symmetric molecules.
For this case, an explicit expression for $V_2^{a b} (k)$ is noted as 
Eqs.~(\ref{eq:m2-exp}) in Appendix~\ref{appen:A}. 

The correlators $\phi_k (t)$ and $F_k^{a b} (t)$ with $a, b = A$
or $B$ are taken from Sec.~\ref{subsec:MCT-v} and Sec.~\ref{subsec:MCT-u}, 
respectively.
The theory of Sec.~\ref{subsec:MCT-C1-MSD} provides the
results for the mean-squared displacement $\delta r_C^2 (t)$. The
Gaussian approximation shall be used to evaluate 
$F_k^{C C} (t) \approx \exp [- \frac{1}{6} k^2 \delta r_C^2 (t)]$. 
The two remaining functions can be expressed in terms of tensor-density
fluctuations according to Eq.~({\ref{eq:Fab-site-tensor}): 
$F_k^{a C} (t) = \sum_\ell
\sqrt{(2 \ell + 1)} j_\ell (k z_a) \phi_{\ell 0} (k 0, t)$
for $a = A$ or $B$. 
As in the previous work \cite{Goetze00c}, only the diagonal correlators
shall be taken in the sum, i.e., 
the approximation will be used: 
$F_k^{a C} (t) \approx j_{0} (k z_{a}) F_k^{C C} (t)$. 

\section{RESULTS}
\label{sec:4}

A few concepts shall be mentioned
which were introduced \cite{Goetze91b} to describe the
MCT-liquid-glass-transition dynamics. In the space of control
parameters, a smooth function $\sigma$ is defined near the
transition points, called the separation parameter. Glass states
are characterized by $\sigma > 0$, liquid states by $\sigma < 0$,
and $\sigma = 0$ defines the transition hypersurface.
Suppose, only one control parameter, say the density $\rho$, is
varied near the transition point. Then one can write for small
distance parameters $\epsilon = (\rho - \rho_c) /\rho_c : \sigma =
C \epsilon$, $C > 0$. In addition to $C$, the transition point is
characterized by a time scale $t_0$ and by a number $\lambda$, 
$0 < \lambda < 1$.
The scale $t_0$ specifies properties of the transient dynamics,
and $\lambda$ is called the exponent parameter. 
The latter determines a certain
number $B > 0$, the critical exponent $a, \, 0 < a \leq 1/2$, and
the von-Schweidler exponent $b, \, 0 < b \leq 1$. There are two
critical time scales governing the 
bifurcation dynamics close to the transition:
\begin{equation}
t_{\sigma} = t_{0} \, / \, | \sigma |^{\delta}, \quad
t'_{\sigma} = t_{0} \, B^{-1/b} \, / \, | \sigma |^{\gamma}.
\label{eq:t-sigma}
\end{equation}
The anomalous exponents of the scales read: $\delta = 1/2 a, \,
\gamma = 1/2a + 1/2 b$. The hard-sphere-system (HSS) shall be
used as solvent. There is only one control parameter for the
equilibrium structure, which shall be chosen as the packing
fraction $\varphi$ of the particles with diameter $d, \, \varphi =
\pi \rho d^3/6$. The distance parameter shall be given by the
logarithm $x$ of $\mid \epsilon \mid$:
\begin{equation}
\epsilon = (\varphi - \varphi_{c}) \, / \, \varphi_{c} = \pm 10^{-x}.
\label{eq:epsilon-def}
\end{equation}
The structure factor $S_q$ is calculated within the Percus-Yevick
theory \cite{Hansen86}. The wave numbers are discretized to $M =
100$ values with spacing $hd = 0.4$. For this solvent model,
results for the density correlators and their
spectra can be found in Ref.~\cite{Goetze00}. The glassy dynamics
is analyzed in Ref.~\cite{Franosch97}, from which one infers: $\varphi_c =
0.516, \, C = 1.54, \, \lambda = 0.735,\, a = 0.312, \, b= 0.583$
and $B = 0.836$. 
Furthermore,  $t_0 = 0.0236 (d/v)$~\cite{Franosch98}.

Dumbbells of two fused hard spheres of diameters $d_A = d_B = d$
shall be used as solute. The elongation parameter 
$\zeta = L / d$ quantifies the bond length. The 
solute-solvent-direct-correlation functions are also
calculated within the Percus-Yevick theory. Within the
tensor-density description, the direct correlation functions
$c_\ell (q)$ have been determined in Ref.~\cite{Franosch97b}.
These results are substituted in the formulas of Appendix A, to
evaluate the equilibrium structure in the site representation. In all
summations over contributions due to various angular-momentum
indices $\ell$, a cutoff $\ell_{co} = 8$ is chosen. It was checked
for representative cases, that increasing the cutoff to $\ell_{co}
= 16$ does not significantly change the results to be discussed.
The discretization of the various wave-vector integrals is done
as specified above for the solvent. 
The results in Secs.~\ref{subsec:result-form-factors} and 
\ref{subsec:result-correlators} deal with a symmetric dumbbell
with $m_A = m_B = m$, and in Sec.~\ref{subsec:result-inertia}, 
the molecule
with $m_A = 10 m$, $m_B = m$ is considered. 

Throughout the rest of this paper, the particle diameter is chosen
as unit of length, $d = 1$, and the unit of time is chosen so that
the thermal velocity of the solvent is $v = 1$.

\subsection{Structural arrest}
\label{subsec:result-form-factors}

There are two control parameters for the system, 
namely the packing fraction $\varphi$ of the solvent and the
elongation $\zeta$ of the solute molecule. Figure~\ref{fig:phase} 
exhibits the phase diagram. 
Phase I deals with states where $\varphi$ is
below the critical value $\varphi_c$, i.e., the solvent is a liquid.
In this case, the long-time limits of the mode-coupling
kernels in Eqs.~(\ref{eq:MCT-u}) vanish. 
All solute correlators relax to zero for long times, 
and the molecule diffuses through the solvent.
For $\varphi \geq \varphi_c$, the solvent is a glass. 
Structural fluctuations behave nonergodically. In particular, a
tagged solvent particle does not diffuse; rather it is localized.
Since the atoms of the molecule with $d_{A} = d_{B} = d$ 
experience the same interaction
with the solvent as the solvent particles among each other, one
expects the molecule to be localized as well. Indeed, Eq.~(\ref{eq:DW-NZ}) 
yields for $\varphi \geq \varphi_c : f_q^N > 0$. If $\varphi$
increases from below $\varphi_c$ to above $\varphi_c$, the
long-time limit $\phi_q^N (t \to \infty)$ increases
discontinuously at $\varphi_c$ from zero to $f_q^{Nc} > 0$. 
Also, the quadrupole correlator exhibits
nonergodic dynamics: $C_2 (t \to \infty) = f_2 > 0$. The cages
surrounding the molecule cause such strong steric hindrance, that
quadrupole fluctuations of the orientational vector $\vec e$
cannot relax to zero. In this sense, the states $\varphi \geq
\varphi_c$ are ideal glasses. 

There are two alternatives for the glass. Phase II deals
with states for sufficiently small $\zeta$. There is
such small steric hindrance for a flip of the molecule's axis
between the two energetically equivalent positions 
$\vec e$ and $- \vec e$ that Eq.~(\ref{eq:DW-NZ}) yields 
$f_q^Z = 0$. The dynamics of the ``charge'' fluctuations is ergodic. In
particular, the dipole correlator relaxes to zero: $C_1 (t \to
\infty) = 0$. Phase II is an amorphous counterpart of a
plastic crystal.
For sufficiently large $\zeta$, steric
hindrance for dipole reorientations is so effective, that also the
``charge'' fluctuations behave nonergodically. In this case, 
Eq.~(\ref{eq:DW-NZ}) yields a positive long-time limit 
$0 < f_q^Z = \phi_q^Z (t \to \infty)$. 
In particular, dipole-disturbances do not relax to
zero: $C_1 (t \to \infty) = f_1 > 0$. This phase III is a glass
with all structural disturbances exhibiting nonergodic motion. 
The two phases II and III are separated by transitions at 
$\zeta = \zeta_c (\varphi)$, $\varphi \geq \varphi_c$. 
With decreasing density,
the steric hindrance for reorientations decreases. Thus, $\zeta_c
(\varphi)$ increases with decreasing $\varphi$, as shown by the
full line in Fig.~\ref{fig:phase}. The transition curve terminates with
horizontal slope at the largest critical elongation  $\zeta_c =
\zeta (\varphi_c) = 0.380$. 
Function $\zeta_{c}(\varphi)$ was
calculated before within the MCT based on the tensor-density
description \cite{Franosch98b}, and the
transition curve of this theory is added as dashed line in 
Fig.~\ref{fig:phase}.
The results of the two theories are in qualitative agreement.
It would be interesting if molecular-dynamics studies would
decide, which of the two theories leads closer to reality.
The asymptotic laws for the transition from phase II to phase III
have earlier been described as type-$A$ transition as
can be inferred from Ref.~\cite{Franosch94} and the papers quoted
there. At this transition, $C_1 (t \to \infty)$
increases continuously with increasing $\zeta$.

The heavy full lines in Fig.~\ref{fig:fq-strong} exhibit critical nonergodicity
parameters $f_q^{x \, c}$ for $\zeta = 0.80$, calculated from 
Eq.~(\ref{eq:DW-NZ})
for the liquid-glass transition point $\varphi = \varphi_c$.
These quantities are Lamb-M\"ossbauer factors of
the molecule. The function $f_{q}^{Nc}$ 
can be measured, in principle, as cross section for incoherent
neutron scattering from the solute, provided both centers $A$ and
$B$ are identical atoms without spin. As expected for a localized
probability-distribution Fourier transform, the
$f_q^{x \, c}$--versus--$q$ curves decrease with increasing $q$. Most
remarkable are the kinks exhibited by $f_q^{Nc}$ for wave numbers $q$
near 5, 12.5 and 20, and by $f_q^{Zc}$ for $q$ near 10 and 17.5.
The light full lines exhibit $f_{q}^{x \, c}$ calculated with 
Eq.~(\ref{eq:Fab-site-tensor-2}) from
the critical nonergodicity parameters $f^c (q \ell 0)$~\cite{Franosch97c}. 
The results of both approximation theories are in semi-quantitative agreement,
in particular concerning the position and size of the kinks. The
$f^c (q \ell 0)$--versus--$q$ curves are bell shaped, close to
Gaussians \cite{Franosch97c}. They enter Eq.~(\ref{eq:Fab-site-tensor-2}) 
with prefactors $j_\ell (q \zeta /2)^2 = O (q^{2 \ell})$, so that the maximum of
the contribution from angular-momentum-index $\ell$ occurs at some 
$q^{\max}_{\ell}$ which increases with $\ell$. The separate contributions
for different $\ell$ are shown as dotted lines in 
Fig.~\ref{fig:fq-strong}. Thus,
the kinks are due to interference effects of the  $f^c (q \ell 0)$
with the intramolecular form factors $j_\ell (q \zeta / 2)$. 
Let us add, that also the Lamb-M\"ossbauer factors of the atoms,
$f_{q}^{a \, c}$, are well described by Gaussians for $q<10$;
in particular these functions do not exhibit kinks.
Figure~\ref{fig:fq-strong} demonstrates for a case of strong steric hindrance for
reorientational motion that angular-momentum variables for $\ell$
up to 6 are relevant to describe the arrested structure, and
that the description of the molecule by site-density
fluctuations properly accounts for the contributions with 
$\ell \geq 2$.

Figure~\ref{fig:fq-weak} exhibits $f_{q}^{x \, c}$
representative for weak
steric hindrance for the reorientational dynamics. Naturally, the
contributions due to the arrest of fluctuations of tensor densities
with large $\ell$ are suppressed. The contributions for $\ell = 0$
and $\ell = 2$ are sufficient to explain $f_q^{Nc}$, in particular
its kink for $q$ near 12.5. Similarly, the contributions for $\ell
= 1$ and $\ell = 3$ are necessary and sufficient to explain
$f_q^{Zc}$ with its kink for $q$ near 17.5. 
The dynamics is strongly influenced by precursor phenomena of the
transition from phase II to phase III. This is demonstrated, for
example, by the strong decrease of $f_1^c = f_{q \to 0}^{Zc}$ for
the result shown in the lower panel of Fig.~\ref{fig:fq-weak} in comparison to the
one shown in the lower panel of Fig.~\ref{fig:fq-strong}. The two
approximation theories under discussion yield different numbers
for the value $\zeta_c$ for the transition point.
It is meaningless to compare different
approximations for results near a singularity $\zeta_c$, referring
to the same value $\zeta$. 
It is more meaningful to compare results for
the same relative distance from the critical point, 
$(\zeta - \zeta_c) /\zeta_c$, as is done in Fig.~\ref{fig:fq-weak}. 
Let us mention that $f_{q}^{Zc}$ shown by the heavy and light full lines would be
somewhat closer, if one had compared elongations yielding the same
value for $f_1^c$.

Figure~\ref{fig:fq-vs-zeta} exhibits critical Lamb-M\"ossbauer factors 
$f_q^{x \, c}$ as function of the
elongation. The lower panel demonstrates the transition from phase
II for $\zeta < \zeta_c$ to phase III for $\zeta > \zeta_c$. 
For strong steric
hindrance, say $\zeta \geq 0.8$, $f_q^{Nc}$ is rather close to
$f_q^{Zc}$ provided $q$ is not too small, say $q > 3$. 
For $\zeta$ approaching $\zeta_c$, the $f_q^{Zc}$ fall below
$f_q^{Nc}$. Most remarkable are the wiggles or even minima of the curves.
These are the analogues of the kinks, discussed above in connection with 
Figs.~\ref{fig:fq-strong} and \ref{fig:fq-weak}. 
Again, these anomalies can be explained as interference
effects between the geometric structure factors 
$j_\ell (q \zeta /2)^2$ and the nonergodicity parameters $f^c (q \ell 0)$ 
according to Eq.~(\ref{eq:Fab-site-tensor-2}). 
Let us consider $f_q^{Nc}$ for
an intermediate wave vector as shown for curves $b$ or $c$ in the
upper panel of Fig.~\ref{fig:fq-vs-zeta}. For small $\zeta$, say $\zeta \leq 0.4$,
the $\ell = 0$ contribution dominates the sum in 
Eq.~(\ref{eq:Fab-site-tensor-2}), as
can be inferred from Fig.~\ref{fig:fq-weak}. Function $f^c (q 0 0)$ reflects the
isotropic part of the arrested fluctuations, and hence it is
practically $\zeta$-independent 
(as shown in Fig.~5 of Ref.~\cite{Franosch97c}). 
Since $j_0 (q \zeta /2)^2 = 1 - \frac{1}{12}
(q \zeta)^2 + O ((q \zeta)^4)$ decreases with increasing $\zeta$,
the $f_q^{Nc}$--versus--$\zeta$ curve decreases too; and the
decrease is stronger for larger $q$. The $f^c (q 2 0)$
increase from 0 for $\zeta = 0$ to values near 0.5 for 
$\zeta = 1$ 
(as shown in Fig.~5 of Ref.~\cite{Franosch97c}). 
Also, the geometric
structure factor increases strongly with $\zeta: j_2 (q \zeta
/2)^2 = ((q \zeta)^2 /60)^2 + O ((q \zeta)^6)$. The combined
effect of both increases causes the increase of the
$f_q^{Nc}$--versus--$\zeta$ curve for larger $\zeta$. The
resulting minimum occurs for smaller $\zeta$ if $q$ is larger, 
and this explains the difference between 
the two curves $b$ and $c$.
The theory produces the minima since it 
accounts for the arrest of tensor-density fluctuations for $\ell
\geq 2$.

\subsection{Correlation functions and spectra near the glass transition}
\label{subsec:result-correlators}

Figure~\ref{fig:NN-ZZ-t} demonstrates the evolution of the dynamics for the
correlators $\phi_q^N (t)$ and $\phi_q^Z (t)$
for intermediate wave numbers $q$ near the transition from phase I
to phase III. The oscillatory transient dynamics occurs within the
short-time window $t < 1$. The control-parameter sensitive
glassy dynamics occurs for longer times for packing
fractions $\varphi$ near $\varphi_c$. At the
transition point $\varphi = \varphi_c$, the correlators decrease
in a stretched manner towards the plateau values $f_{q}^{x \, c}$ 
as shown by the dotted lines. Increasing $\varphi$
above $\varphi_c$, the long-time limits increase, as shown for
$\phi_q^Z (t)$ for $q = 7.4$ and $\zeta = 0.80$. Decreasing
$\varphi$ below $\varphi_c$, the correlators cross the plateau at
some time $\tau_\beta$, and then decay towards zero. The
decay from the plateau $f_q^{x \, c}$ to zero is the $\alpha$-process
for $\phi_q^x (t)$. It is characterized by a time scale
$\tau_\alpha$, which can be defined, e.g., by 
$\phi_q^x (\tau_\alpha) = f_q^{x \, c} / 2$. 
Upon decreasing $\varphi_c - \varphi$ towards zero, 
the time scales $\tau_\beta$ and
$\tau_\alpha$ increase towards infinity proportional to $t_\sigma$
and $t_\sigma^\prime$, respectively, cited in 
Eq.~(\ref{eq:t-sigma}). The figure
exemplifies the standard MCT-bifurcation scenario. For small
$\mid \varphi - \varphi_c \mid$, the results can be described
in terms of scaling laws. This was explained in
Refs.~\cite{Franosch97,Fuchs98} for the HSS, 
and the discussion shall not be
repeated here. 

One can deduce from Fig.~\ref{fig:fq-strong} 
that for $\zeta = 0.80$ and $q \geq 5$
the plateaus for both types of density fluctuations are very close
to each other: $f_q^{Nc} \approx f_q^{Zc}$. The upper two panels
of Fig.~\ref{fig:NN-ZZ-t} demonstrate that also the dynamics is nearly the same,
$\phi_q^{N} (t) \approx \phi_q^{Z} (t)$. This means that
for $q \zeta > 4$ and for strong steric hindrance, 
the cross correlations $F_q^{AB} (t)$ are
very small. The reason is that the
intramolecular correlation factors $j_\ell (q \zeta /2)$ are
small, and thus interference effects between the density
fluctuations of the two interaction sites are suppressed.
Coherence effects can be expected only for smaller wave vectors.
For this case, the functions can be understood in terms of
their small-$q$ asymptotes, 
Eq.~(\ref{eq:Fab-small-q}). 

The lower two panels in Fig.~\ref{fig:NN-ZZ-t} deal with weak steric hindrance.
In this case, the ``charge''-density
fluctuations behave quite differently from the number-density
fluctuations. The most important origin of this difference is the
reduction of the mode-coupling vertices $V_{q,kp}^Z$ relative to
$V_{q,kp}^N$ in Eq.~(\ref{eq:MCT-NZ-b}). For small elongations of the molecule,
the effective solute-solvent potentials for reorientations are
small. Therefore, the $f_q^{Zc}$ decrease strongly relative to
$f_q^{Nc}$ for $\zeta$ decreasing towards $\zeta_c$, as is shown
in Fig.~\ref{fig:fq-vs-zeta}. 
Upon approaching $\zeta_{c}$, 
the $\alpha$-peak strength of $\phi_q^Z (t)$ gets suppressed
relative to the one of $\phi_q^N (t)$. Within phase II, 
the ``charge''-density fluctuations
relax to zero as in a normal liquid.
This implies as precursor phenomenon, that
the time scale $\tau_\alpha^Z$ of the
``charge''-density-fluctuation $\alpha$-process 
shortens relative to the scale $\tau_\alpha^N$ for the
number-density fluctuations. 
Thus, the small-$\zeta$ behavior shown in the lower two panels
of Fig.~\ref{fig:NN-ZZ-t} is due to disturbances of the standard MCT-transition
scenario by the nearby type-$A$ transition.

The correlators $C_1 (t)$ and $C_2 (t)$ are
shown in Fig.~\ref{fig:C1-C2} for the critical point $\varphi = \varphi_c$ and
for two liquid states near the transition from phase I to phase
III. For $\zeta = 0.80$, the anisotropic
distribution of the solvent particles around the molecule leads to
a stronger coupling to the dipole reorientations than to the
reorientations for the quadrupole, and therefore the plateau for
the former is higher than for the latter, $f_1^c > f_2^c$. A
leading order expansion of the solutions of the equations of
motion (\ref{eq:GLE-C1}) and (\ref{eq:GLE-C2}) 
in terms of the small parameter $C_\ell (t) - f_\ell^c$ 
leads to the factorization in the critical amplitude
$h_\ell$ and a function $G (t)$ called the $\beta$-correlator,
$C_{\ell}(t) - f_{\ell}^{c} = h_{\ell} G(t)$.
The latter is the same for all correlation functions. It obeys the
first scaling law of MCT,
$G(t) = \sqrt{| \sigma |} \, g_{\pm} (t / t_{\sigma})$ for
$\sigma \gtrless 0$.
The master functions $g_{\pm} (\hat t)$ are determined 
by $\lambda$. 
They also describe the dynamics of the
solvent in the window where 
$\mid \phi_q (t) - f_q^c \mid \ll 1$~\cite{Franosch97}. In particular
there holds $g_- (\hat t_-) = 0, \, \hat t_- = 0.704$, so that
both correlators $C_\ell (t)$ cross their plateau at the same time
$\tau_\beta = \hat t_- t_\sigma$. The non-linear mode-coupling
effects request, that the correlators approach zero roughly at the
same time. Thus one understands the general differences between the
$\alpha$-processes, which were mentioned in the introduction:
the $\alpha$-process for dipole relaxation is stronger, slower,
and less stretched than the one for quadrupole relaxation. This
finding is in qualitative agreement with the ones of
the theory based on tensor-density representation of the
structure \cite{Goetze00c}. There are, however, quantitative
differences between the two approximation schemes. The plateaus
$f_1^c = 0.905$ and $f_2^c = 0.674$ are smaller
than the corresponding values 0.943 and 0.835 found in 
Ref.~\cite{Goetze00c} and the amplitudes $h_1 = 0.19$ and $h_2 = 0.40$
are bigger than the corresponding values 0.13 and 0.35 calculated
previously \cite{Goetze00c}. The times $\tau_\alpha^\ell$
characterizing the $\alpha$-process shall be defined by $C_\ell
(\tau_\alpha^\ell) = f_\ell^c / 2$. They are marked by open
squares in Fig.~\ref{fig:C1-C2}. 
The values $\tau_{\alpha}^{1}=5.21 \times 10^{5}$, 
$\tau_{\alpha}^{2}=1.64 \times 10^{5}$ for $x=3$ and $\zeta=0.80$
are smaller than the ones
reported in Ref.~\cite{Goetze00c}. 
The present theory implies a somewhat weaker coupling of the
reorientational degrees of freedom of the molecule to the
dynamics of the solvent, than found earlier \cite{Goetze00c}. 
This holds also for the small elongation $\zeta = 0.43$. The
approach towards the transition from phase III to phase II, leads
to a suppression of $f_1^c$, as discussed for the $f_q^{Zc}$ in
Fig.~\ref{fig:fq-vs-zeta}. 
The dipole relaxation speeds up for $\zeta \to \zeta_c$,
as discussed for the lower panels of Fig.~\ref{fig:NN-ZZ-t}.
This is reflected by an enhancement of $h_1 = 1.60$ relative to the
amplitudes cited for $\zeta = 0.80$ but also relative to 
$h_2 = 0.49$. 

One can perform 
$\lim_{\sigma \to 0-} \phi_q (\tilde{t} t_\sigma^\prime) = 
\tilde{\phi}_q (\tilde{t})$
for the solutions of Eq.~(\ref{eq:GLE-v}), 
where $\tilde{\phi}_q (\tilde{t})$ can be evaluated from 
the mode-coupling functional at the critical point. It obeys the
initial condition 
$\tilde \phi_q (\tilde t) = f_q^c - h_q \tilde{t}^b + O (\tilde t^{2b})$. 
Function $\tilde \phi_q (\tilde t)$ can
be considered as shape function of the $\alpha$-process, and the
result implies the second scaling law of MCT, also referred to as
superposition principle: $\phi_q (t) = \tilde \phi_q (t /
t_\sigma^\prime)$ for $\sigma \to 0-$ \cite{Goetze91b}.
Corresponding laws hold for all functions, as is demonstrated in
detail for the HSS in Refs.~\cite{Franosch97,Fuchs98}. 
In particular, one gets for the reorientational correlators
for $\sigma \to 0-$:
\begin{mathletters}
\label{eq:alpha}
\begin{equation}
C_{\ell}(t) = \tilde{C}_{\ell}(\tilde{t}), \quad
\tilde{t} = t / t'_{\sigma}, \quad
t_{\sigma} \ll t,
\label{eq:alpha-a}
\end{equation}
and this corresponds to the $\alpha$-scaling law for the susceptibility
spectra
\begin{equation}
\chi''_{\ell}(\omega) = \tilde{\chi}''_{\ell}(\tilde{\omega}), \quad
\tilde{\omega} = \omega t'_{\sigma}, \quad
\omega \ll 1/t_{\sigma}. 
\label{eq:alpha-b}
\end{equation}
The initial decay of the master function $\tilde{C}_{\ell}(\tilde{t})$ 
is described by von
Schweidler's law
\end{mathletters}
\begin{mathletters}
\label{eq:von}
\begin{equation}
\tilde{C}_{\ell}(\tilde{t}) = f_{\ell}^{c} - h_{\ell} \, \tilde{t}^{b}, \quad
\tilde{t} \to 0,
\label{eq:von-a} 
\end{equation}
which is equivalent to a power-law tail of the master spectrum
$\tilde{\chi}''_{\ell}(\tilde{\omega})$:
\begin{equation}
\tilde{\chi}''_{\ell}(\tilde{\omega}) = 
h_{\ell} \, \sin (\pi b / 2) \, \Gamma(1+b) \, / \, \tilde{\omega}^{b}, \quad
\tilde{\omega} \to \infty. 
\label{eq:von-b}
\end{equation}
The upper panel of Fig.~\ref{fig:alpha-spectra} 
exhibits the $\alpha$-process master spectra for the
reorientational processes for $\zeta = 0.80$ and 
for the dimensionless longitudinal 
elastic modulus $m_{q=0}(t)$ of the solvent.
The latter can be measured by Brillouin-scattering spectroscopy.
It probes a tensor-density fluctuation for 
$\ell = 0$. 
\end{mathletters}
The von-Schweidler-law tails describe the spectra
for frequencies exceeding the position $\tilde \omega_{\max}$
of the susceptibility maximum by a factor of about 100, as shown
by the dashed lines. Since $f_1^c > f_2^c$ and both plateau values are rather
large, one understands from the theory for the leading corrections
to Eq.~(\ref{eq:von-b}) \cite{Franosch97} that for decreasing $\tilde \omega$
the von-Schweidler asymptote underestimates the spectrum, and this
by larger values for $\ell = 1$ than for $\ell = 2$. 
For smaller plateau values, the von-Schweidler asymptote may
overestimate the spectrum, as is exemplified for the modulus.
The lower panel of Fig.~\ref{fig:alpha-spectra}
demonstrates that the $\alpha$-processes speed up if steric
hindrance is decreased. As precursor of the transition to phase II,
the spectrum for the dipole response is located at much higher frequencies
than that for the quadrupole response. 
Traditionally, dielectric-loss spectra have been fitted by the
ones of the Kohlrausch-law 
$\phi_K (\tilde{t}) = A \exp[ - (\tilde{t} B)^\beta]$. 
Such fits also describe a major part of the
spectra in Fig.~\ref{fig:alpha-spectra}, as shown by the dotted lines. The parameters
$A$ and $B$ are adjusted to match the susceptibility maximum. The
stretching exponent $\beta$ is chosen so that the spectrum is
fitted at half maximum $\tilde\chi_{\ell}^{\prime \prime} (\tilde
\omega_{\max}) / 2$. If one denotes the width in $\log_{10}
\omega$ at half maximum by $W$, stretching means that this
parameter is larger than the value $W_D = 1.14$ characterizing a
Debye process, $\phi_D (\tilde t) = \exp( - \tilde{t})$. 
The upper panel of Fig.~\ref{fig:alpha-spectra}
quantifies the general results of the theory for strong steric
hindrance: $\tilde \chi_1^{\prime \prime} (\tilde \omega_{\max}^1)
> \tilde \chi_2^{\prime \prime} (\tilde \omega_{\max}^2)$,
$\tilde \omega_{\max}^1 < \tilde \omega_{\max}^2$ and $\beta_1 >
\beta_2$. It quantifies also the forth property cited in the
introduction $\tilde \omega_{\max}^2 < \tilde \omega_{\max}^0$. 
A further general property is $\beta_2 > \beta_0$.

The discussion of power-law spectra is done more
conveniently in a double logarithmic diagram
as shown in Fig.~\ref{fig:Nagel-plot} for normalized
dipole-fluctuation-$\alpha$-process spectra 
$\tilde{C}_{1}^{\prime \prime} (\tilde{\omega}) \tilde{\omega}_{\max} / f_{1}^{c} =
\tilde{\chi}_1^{\prime \prime} (\tilde{\omega}) \tilde{\omega}_{\max} /
f_{1}^{c} \tilde{\omega}$ as function of 
$\tilde{\omega} / \tilde{\omega}_{\max}$.
One notices, that there is a
white-noise spectrum for $\tilde{\omega}$ below 
$\tilde{\omega}_{\max}$. The high-frequency
wing of the Kohlrausch-law fit decreases proportional to
$\tilde{\omega}^{- \beta}$ and underestimates the
spectrum $\tilde\chi_1^{\prime \prime}(\tilde{\omega})$ considerably. Because of
the von-Schweidler asymptote, which is shown as dashed straight
line, the spectrum exhibits an enhanced high-frequency wing.
Dixon {\it et al.} \cite{Dixon90} made the remarkable observation that 
their dielectric spectra could be
collapsed on one master curve, if the vertical axis is rescaled by
$w^{-1}$ and the horizontal axis by $w^{-1} (1 + w^{-1})$, where 
$w = W / W_D$.
In Fig.~\ref{fig:Nagel-plot} this scaling is used and the data for
glycerol from Ref.~\cite{Dixon90} are included. The spectra 
for molecules with $\zeta = 0.80$ and $\zeta=0.60$,
which are relevant for the description of van der Waals 
systems~\cite{Goetze00c}, follow the
mentioned scaling surprisingly well.
This finding appears non-trivial, since the scaling is not reproduced
by the MCT results of the basic quantities $\phi_{q}(t)$~\cite{Fuchs92b}.
The rescaled spectrum for $\zeta=0.43$ deviates from the scaling law
for $w^{-1}(1+w^{-1}) \log_{10} (\tilde{\omega}/\tilde{\omega}_{max}) \geq 5$.

It might appear problematic that the dipole correlator was
calculated within a different approximation scheme than the
quadrupole correlator. But, there is no difficulty to also
evaluate $C_1 (t)$ within the scheme explained in 
Sec.~\ref{subsec:MCT-C2} for
the evaluation of $C_2 (t)$. Figure~\ref{fig:compare-C1} 
presents a comparison of 
$C_1 (t)$ obtained along the two specified routes. 
The two results for the small elongation $\zeta =
0.43$ are close to each other. The discrepancies are mainly due to the
7\% difference between the two plateau values $f_1^c$.
With increasing $\zeta$, the discrepancies
decrease. For the large elongation $\zeta = 0.80$, the results are
practically undistinguishable.

\subsection{Structural relaxation versus transient dynamics}
\label{subsec:result-inertia}

Let us introduce Fourier-Laplace transforms of functions of time,
say $f (t)$, to functions of frequency, say $f (\omega)$, with the
convention $f (\omega) = i \int_0^\infty dt \exp (izt) f (t)$, 
$z = \omega + i0$. 
The equations of motion (\ref{eq:GLE-u}) with the
initial conditions from Eq.~(\ref{eq:Fab-short-time}) are transformed to
\begin{equation}
[\omega \, {\bf 1} + {\bf \Omega}_{q}^{2} \, {\bf m}_{q}(\omega)] \,
[\omega \, {\bf F}_{q}(\omega) + {\bf w}_{q}] -
{\bf \Omega}_{q}^{2} \, {\bf F}_{q}(\omega) = {\bf 0}.
\label{eq:str-1}
\end{equation}
Within the glass, the long-time limits of ${\bf F}_q (t)$ and
${\bf m}_{q}(t)$ do not vanish, i.e., the transformed quantities exhibit
zero-frequency poles. One gets, for example, $\omega {\bf F}_q
(\omega) \to - {\bf F}_q^\infty$ for $\omega \to 0$, where the
strength $- {\bf F}_q^\infty$ 
of the poles follow from Eq.~(\ref{eq:DW-u}). Continuity of the
solutions of the MCT-equation of motion implies that for vanishing
frequencies and for vanishing distances from the transition points,
${\bf m}_q (\omega)$ becomes arbitrarily large. Hence, combinations like 
$\omega + i \xi_q$ with constants $\xi_{q}$
can be neglected compared to ${\bf m}_q (\omega)$.
Therefore, in the region of glassy dynamics, Eq.~(\ref{eq:str-1}) can be
modified to
\begin{equation}
{\bf F}_{q}(\omega) - {\bf m}_{q}(\omega) \, {\bf w}_{q} =
i \,\mbox{\boldmath $\xi$}_{q} +
\omega \, {\bf m}_{q}(\omega) \, {\bf F}_{q}(\omega).
\label{eq:str-2}
\end{equation}
Let us assume, that this equation has a solution, to be denoted by
${\bf F}_q^{*} (\omega)$, which is defined for all frequencies,
so that it can be back
transformed to a function ${\bf F}_q^{*} (t)$, defined for all $t > 0$. 
Choosing $\mbox{\boldmath $\xi$}_{q}$
properly, Eq.~(\ref{eq:str-2}) can
be written as 
\begin{mathletters}
\label{eq:str-3}
\begin{equation}
{\bf F}_{q}^{*}(t) - {\bf m}_{q}^{*}(t) \, {\bf w}_{q} =
- (d/dt) \int_{0}^{t} dt' \, {\bf m}_{q}^{*}(t-t') \, {\bf F}_{q}^{*}(t').
\label{eq:str-3-a}
\end{equation}
Similar reasoning leads from Eq.~(\ref{eq:GLE-v}) to
\begin{equation}
\phi_{q}^{*}(t) - m_{q}^{*}(t) = - (d/dt) \int_{0}^{t} dt' \,
m_{q}^{*}(t-t') \, \phi_{q}^{*}(t').
\label{eq:str-3-b}
\end{equation}
These formulas have to be supplemented with the MCT expressions
for the kernels
\begin{equation}
{\bf m}_{q}^{*}(t) = \mbox{\boldmath ${\cal F}$}_{q}
[ {\bf F}^{*}(t), \phi^{*}(t) ], \quad
m_{q}^{*}(t) = {\cal F}_{q}[\phi^{*}(t)].
\label{eq:str-3-c}
\end{equation}
Equations (\ref{eq:str-3}) for the glassy dynamics are scale
invariant. 
\end{mathletters}
With one set of solutions $\phi_q^{*} (t)$ and ${\bf
F}_q^{*} (t)$, also the set 
$\phi_q^{* \, u} (t) = \phi_q^{*} (ut)$ and 
${\bf F}_q^{* \, u} (t) = {\bf F}_q^{*} (ut)$
provides a solution for arbitrary $u > 0$. 
To fix the solution uniquely, one can 
introduce positive numbers $y_q$ and positive definite matrices
${\bf y}_q$ to specify the initial condition as
power-law asymptotes~\cite{Fuchs99b}
\begin{equation}
{\bf F}_{q}^{*}(t) \, (t/t_{0})^{1/3} \to {\bf y}_{q}, \,\,
\phi_{q}^{*}(t) \, (t/t_{0})^{1/3} \to y_{q}, \,\,
(t/t_{0}) \to 0.
\label{eq:str-4}
\end{equation}

The theory of the asymptotic solution of the MCT equations for
simple systems had been built on the analogue of Eq.~(\ref{eq:str-2}) with
$\xi_q$ neglected \cite{Franosch97}. The present theory extends
the previous one by the introduction of matrices. It seems obvious
that the previous results for asymptotic expansions hold in a
properly extended version. Let us only note 
the formula for the long-time decay of the correlators at
the critical point $\varphi = \varphi_c$
\cite{Franosch97,Fuchs98}:
\begin{equation}
{\bf F}_{q}(t) = {\bf F}_{q}^{\infty \, c} +
{\bf H}_{q} (t_{0}/t)^{a}
[{\bf 1} + {\bf K}_{q} (t_{0}/t)^{a} + {\bf O}( (t_{0}/t)^{2a} )].
\label{eq:str-5}
\end{equation}
The exponent $a$, mentioned above, can be calculated from
the mode-coupling functional at the critical point. The same is
true for the plateau values ${\bf F}_q^{\infty \, c}$, and the 
amplitudes ${\bf H}_q$ and ${\bf K}_q$. 
The dependence of the solution from the transient
dynamics is given by the single number $t_0$. 
Let us anticipate that
Eq.~(\ref{eq:str-5}) and similar results can be extended to a complete
solution. One concludes that the glassy dynamics is determined, up
to a scale $t_0$, by the mode-coupling functionals in 
Eq.~(\ref{eq:str-3-c}).

Equations (\ref{eq:BXX-1}) and (\ref{eq:BXX-2}) show, that the mode-coupling
functionals ${\cal F}_q$ and ${\cal F}_{q}^{ab}$ 
are specified by the
density $\rho$, the static structure factor $S_q$, the direct
correlation function $c_q$ of the solvent, and the
solute-solvent direct correlation functions $c_q^a$,
i.e., by equilibrium quantities. They are the same for
systems with a Newtonian dynamics, as considered in this paper,
and for a model with a Brownian dynamics, as is to be used for the
description of colloidal suspensions. In particular, the 
mode-coupling functionals are independent of the particle masses $m,
m_A$ and $m_B$. Thus, the glassy dynamics of the
molecule in the simple liquid does not depend on the 
inertia parameters specifying the microscopic equations of motion.
The same conclusions on the glassy dynamics, which were cited in
the introduction for the basic version of MCT, hold for the model
studied in this paper. Let us add, that neither the temperature
$T$, nor the interparticle-interaction potentials $V$ of the
solvent, nor the solute-solvent-interaction potentials $V^a$
enter explicitly the mode-coupling functionals. These quantities
only enter implicitly via $S_{q}$, $c_{q}$, 
$c_{q}^{A}$ and $c_{q}^{B}$.

The independence of the glassy dynamics on the
inertia parameters is demonstrated in Fig.~\ref{fig:C1-inertia} 
for four states of
the liquid. It is shown that the reorientational dynamics of the
dipole does not change for $t > 1$ even if the mass ratio of the atoms
$m_A /m_B$ is altered by a factor 10. The transient
dynamics, which deals with overdamped librations, 
exhibits an isotope effect. There is
no fitting parameter involved in the shown diagram. The 
scale $t_0$ neither depends on the density of the
solvent nor on the elongation parameter, 
and Eqs.~(\ref{eq:str-3}) describe the complete 
control-parameter dependence of the glassy dynamics.

\section{SUMMARY}
\label{sec:5}

The MCT for simple systems with a dilute solute
of atoms has been generalized to the one with a dilute solute of
diatomic molecules. The derived equations of motion generalize
the ones for atoms in the sense that scalar functions
are replaced by 2-by-2 matrix functions. 
These generalizations result
from the description of the position of the molecule in
terms of interaction-site-density fluctuations. The numerical
effort required for a solution of the equations is not seriously
larger than the one needed to solve the corresponding equations
for atomic solutes. This holds in particular for symmetric
molecules where the matrix equations can be diagonalized
by a linear transformation to number-density and
``charge''-density fluctuations. 
The theory implies
that the dynamics outside the transient regime is determined, up
to an overall time scale $t_0$, by the equilibrium structure. In
particular, it is independent of the inertia
parameters of the system.

It was shown that the present theory reproduces all qualitative
results obtained within the preceding much more involved theory based
on a description of the solute by tensor-density fluctuations.
Both theories yield a similar phase diagram, Fig.~\ref{fig:phase}. 
The characteristic differences of the reorientational correlators
exhibited for strong steric hindrance of rotations as opposed to
weak steric hindrance are obtained here, Fig.~\ref{fig:C1-C2}, 
as previously.
There are systematic quantitative differences between the two
approximation approaches in the sense, that the implications of
the cage effect are weaker in the present theory than in the
preceding work \cite{Franosch97c,Goetze00c}. The earlier findings
on the differences of the spectra referring to
responses with angular-momentum index $\ell = 0$, 1 and 2 have
been corroborated by separating the $\alpha$-peaks from
the complete spectra with the aid of the $\alpha$-scaling law,
Fig.~\ref{fig:alpha-spectra}. 

The critical non-ergodicity parameters $f_q^{x \, c}$ have been
calculated, which determine the form factors of
quasi-elastic scattering from the liquid near the glass transition.
Contrary to what one finds for atomic solutes, these are not
bell-shaped functions of wave numbers $q$, rather they exhibit
kinks, Fig.~\ref{fig:fq-strong}. The form factors for some 
wave numbers $q$ 
vary non-monotonically with changes of the elongation of the molecule, 
Fig.~\ref{fig:fq-vs-zeta}.
Analyzing these findings in terms of form factors defined for
fixed angular-momentum index $\ell$, one can explain the
results as due to intra-molecular interference effects 
demonstrating that the theory accounts for reorientational
correlations with angular-momentum index $\ell \geq 2$.

It was shown that the $\alpha$-relaxation spectra for dipole
reorientations with strong steric hindrance obey the scaling law
proposed by Dixon {\it et al.} \cite{Dixon90} within the window and
within the accuracy level considered by these authors. There is no
fitting parameter involved in the construction of 
Fig.~\ref{fig:Nagel-plot}, which
demonstrates this finding for the two elongation parameters
$\zeta=0.60$ and 0.80. Thus, it is not
justified to use the cited empirical scaling as an argument against the
applicability of MCT for a discussion of dielectric-loss spectra.

\acknowledgments

We thank M.~Fuchs, M.~Sperl and Th.~Voigtmann for many
helpful discussions and suggestions. 
We are grateful to A.~Latz and R.~Schilling for constructive
critique of our manuscript.
S.-H.~C. acknowledges financial support from
JSPS Postdoctoral Fellowships for Research Abroad. 
This work was supported by 
Verbundprojekt BMBF 03-G05TUM.

\appendix

\section{TENSOR-DENSITY REPRESENTATIONS}
\label{appen:A}

Following the conventions of Refs.~\cite{Franosch97c} and
\cite{Goetze00c}, normalized tensor-density fluctuations shall be used by
decomposing the molecule's position variable in plane waves 
$\exp( i {\vec q} \cdot {\vec r}_{C})$ for the center of mass $\vec r_C$ 
and in spherical harmonics $Y_{\ell}^{m}({\vec e} \,)$ for the 
orientation vector $\vec e$:
\begin{equation}
\rho_{\ell}^{m}({\vec q}) = i^{\ell} \, \sqrt{4 \pi} \, 
\exp ( i {\vec q} \cdot {\vec r}_{\rm C}) \,
Y_{\ell}^{m}({\vec e}).
\label{eq:rho-tensor-def}
\end{equation}
The molecule-solvent interactions are specified by a set of
direct correlation functions
\begin{equation}
c_{\ell}(q) = 
\langle \rho_{{\vec q}_{0}}^{*} \, \rho_{\ell}^{0}({\vec q}_{0}) \rangle \, / \, 
(\rho S_{q}); \quad
{\vec q}_{0} = (0,0,q).
\label{eq:cuv-tensor-def}
\end{equation}
The structure dynamics is described by the matrix of correlators,
defined by
\begin{equation}
\phi_{\ell k}(qm,t) = \langle \rho_{\ell}^{m}({\vec q}_{0},t)^{*} 
\rho_{k}^{m}({\vec q}_{0}) \rangle.
\label{eq:tensor-correlator}
\end{equation}

Since the position vectors of the interaction sites can be written as 
$\vec r_a = \vec r_C + z_a \vec e$, 
the Rayleigh expansion of the exponential in Eq.~(\ref{eq:rho-def}) 
yields the formula
\begin{equation}
\rho^{a}_{{\vec q}_{0}} = \sum_{\ell} \sqrt{2 \ell + 1} \, 
j_{\ell}(q z_{a}) \, \rho_{\ell}^{0}({\vec q}_{0}).
\label{eq:rho-site-tensor}
\end{equation}
Substitution of this expression into Eq.~(\ref{eq:huv-def}) leads with 
Eq.~(\ref{eq:cuv-tensor-def})
to an expression connecting the direct correlation functions
$c_q^a$ with $c_{\ell}(q)$:
\begin{equation}
\sum_{b} w^{ab}_{q} \, c^{b}_{q} = 
\sum_{\ell} \sqrt{2 \ell + 1} \, j_{\ell}(q z_{a}) \, c_{\ell}(q).
\label{eq:cuv-site-tensor}
\end{equation}
Substitution of Eq.~(\ref{eq:rho-site-tensor}) into Eq.~(\ref{eq:Fab-def}) 
leads with Eq.~(\ref{eq:tensor-correlator}) to the
expression of the site-density correlators in terms of the
tensor-density correlators:
\begin{equation}
F^{ab}_{q}(t) = \sum_{\ell k} \sqrt{(2 \ell + 1)(2 k + 1)} \, 
j_{\ell}(qz_{a}) \, j_{k}(qz_{b}) \,
\phi_{\ell k}(q0,t).
\label{eq:Fab-site-tensor}
\end{equation}
For the long-time limits of the correlators $F_q^{ab}(t)$ one gets a
combination of the non-ergodicity parameters $f_{\ell k} (q m) =
\phi_{\ell k} (q m, t \to \infty)$. If one uses the diagonal
approximation $f_{\ell k} (q, m) = \delta_{\ell k} f(q \ell m)$,
one can write
\begin{equation}
F^{ab}_{q}(t \to \infty) = \sum_{\ell} ( 2 \ell + 1) \, 
j_{\ell}(qz_{a}) \, j_{\ell}(qz_{b}) \,
f(q \ell 0).
\label{eq:Fab-site-tensor-2}
\end{equation}

Substituting the Rayleigh expansion of Eq.~(\ref{eq:rho-def}) into 
Eq.~(\ref{eq:m2-MCT-a}),
one can express the pair modes in terms of those formed with
tensor-density fluctuations:
\begin{equation}
A^{a}({\vec k}, {\vec p} \, ) = \sqrt{4 \pi} \sum_{\ell m}
j_{\ell}(k z_{a}) \, Y_{\ell}^{m}({\vec k}) \,
[\, \rho_{\ell}^{m}({\vec k}) \, \rho_{\vec p} \, / \sqrt{N S_{p}} \, ].
\end{equation}
Therefore, the overlaps of the forces with the pair modes can be
expressed as sums of the corresponding quantities calculated in
Ref. \cite{Franosch97c}. One finds for the mode-coupling
coefficients in Eq.~(\ref{eq:m2-final}):
\begin{mathletters}
\label{eq:m2-exp}
\begin{equation}
V_{2}^{ab}(k) = [ k^{2} \rho S_{k} / 2 \pi^{2}] 
\sum_{c d} ({\bf w}_{k}^{-1})^{ac} \, D_{c}(k) \, D_{d}(k) \,
({\bf w}_{k}^{-1})^{db}.
\label{eq:m2-exp-a}
\end{equation}
Here one gets for $a = A, B$ or $C$
\bea
D_{a}(k) &=& 
{\textstyle \frac{1}{12}} \sum_{\ell J}
(-1)^{\frac{1}{2} (\ell + J)} \, (2 \ell + 1) \, \sqrt{2J + 1} \,
j_{\ell}(k z_{a}) \, c_{J}(k) 
\nonumber \\
&\times&
[ \, J ( J + 1 ) + 6 - \ell ( \ell + 1 ) \, ]
\left(
\begin{array}{ccc}
2 & \ell & J \\
0 & 0    & 0 
\end{array}
\right)^{2},
\label{eq:m2-exp-b}
\eea
where the last factor denotes Wigner's 3-$j$ symbol. 
\end{mathletters}

\section{Mode-Coupling Coefficients}
\label{appen:B}

Mode-coupling equations based on a description of the molecules
by site-density fluctuations have been derived in Ref.
\cite{Chong98b} by extending the procedure used originally for
atomic systems \cite{Goetze91b}. But the reported formulas
\cite{Chong98b} do not seem appropriate, since they do not reduce
to the ones for tagged particle motion if the limit of a vanishing
elongation parameter $\zeta$ is considered. Therefore, an
alternative derivation will be presented which starts from the
theory developed by Mori and Fujisaka \cite{Mori73} for an
approximate treatment of nonlinear fluctuations. The application
of this theory will be explained by rederiving the equations for
the solvent before the ones for the solute are worked out.

\subsection{The Mori-Fujisaka equations}
\label{subsec:B1}

Let us consider a set of distinguished dynamical variables
$A_\alpha, \alpha = 1,2, \cdots$, defined as functions on the
system's phase space. The time evolution is generated by the
Liouvillian, $A_\alpha (t) = \exp (i {\cal L} t) A_\alpha$. Using
the canonical averaging to define scalar products in the space of
variables, $(A, B) = \langle A^* B \rangle$, the Liouvillian is
Hermitian. It is the goal to derive equations of motion for the
matrix of correlators $C_{\alpha\beta} (t) = (A_\alpha (t),
A_\beta)$. The initial condition is given by the positive definite
matrix $g_{\alpha\beta} = (A_\alpha, A_\beta)$. The theory starts
with a generalized Fokker-Planck equation for the distribution
function $g_a = \Pi_\alpha \, \delta (A_\alpha - a_\alpha), \, a =
(a_1, a_2, \cdots)$. It is assumed that the time scales for the
fluctuations of the $A_\alpha$ and their products is larger than
the ones for the Langevin fluctuating forces. The spectra of the
latter can then be approximated by a constant matrix
$\Gamma_{\alpha\beta}$. It is assumed furthermore, that the
$\Gamma_{\alpha\beta}$ are independent of the distinguished
variables, and that the equilibrium distribution of the latter is
Gaussian:
\begin{equation}
\langle g_{a} \rangle = C \,
\exp \biggl( - \frac{1}{2} \sum_{\alpha \beta}
a_{\alpha} \, g_{\alpha \beta}^{-1} \, a_{\beta}^{*} \biggr).
\label{eq:B1}
\end{equation}
In the cited Fokker-Planck equation there occurs the 
streaming velocity $v_\alpha (a)$ given by
\begin{equation}
v_{\alpha}(a) \, \langle g_{a} \rangle = 
\langle \dot{A}_{\alpha}^{*} \, g_{a} \rangle =
k_{B}T \sum_{\beta} \frac{\partial}{\partial a_{\beta}}
\langle \{ A_{\alpha}^{*}, A_{\beta} \} \, g_{a} \rangle.
\label{eq:B2}
\end{equation}

The Fokker-Planck equation is now reduced by projecting out the subspace
of distinguished variables. There appears the frequency matrix
specifying the linear contribution to the streaming term,
\begin{equation}
\Omega_{\alpha \beta} = \sum_{\gamma}
(A_{\alpha}, {\cal L} A_{\gamma}) \, g_{\gamma \beta}^{-1}.
\label{eq:B3}
\end{equation}
The nonlinear contributions enter as combination
\begin{equation}
f_{\alpha} = \int da \, v_{\alpha}(a) \, g_{a} -
i \sum_{\beta} \Omega_{\alpha \beta} A_{\beta}.
\label{eq:B4}
\end{equation}
They determine the relaxation kernel as
\begin{equation}
M_{\alpha \beta}(t) = \sum_{\gamma} (f_{\alpha}(t), f_{\gamma}) \,
g_{\gamma \beta}^{-1}.
\label{eq:B5}
\end{equation}
The time evolution is generated by the reduced Liouvillian ${\cal
L}^\prime = {\cal Q L Q}, \, f_\alpha (t) = \exp (i {\cal
L}^\prime t) f_\alpha$, where $\cal Q$ is the projector on the
space perpendicular to the one spanned by the distinguished
variables. The result, which is equivalent to 
Eq.~(2.15) in Ref.~\cite{Mori73}, reads
\bea
\partial_{t} C_{\alpha \beta}(t) &=& 
- \sum_{\gamma} [ 
(i \Omega_{\alpha \gamma} + \Gamma_{\alpha \gamma}) \, C_{\gamma \beta}(t) 
\nonumber \\
& &
\quad \quad
+ \,
\int_{0}^{t} dt' \, M_{\alpha \gamma}(t-t') \, C_{\gamma \beta}(t')].
\label{eq:B6}
\eea

\subsection{MCT equations for simple systems}

To get a description of slowly varying structural fluctuations, the
original reasoning of MCT shall be adopted, and as distinguished
variables the density fluctuations $\rho_{\vec q}$ and the
longitudinal current fluctuations 
$j_{\vec q} = \sum_{\kappa} v_{\kappa,z} \exp(i {\vec q} \cdot {\vec r}_{\kappa})$
will be chosen. The variable label of the
preceding subsection consists of two bits, 
$\alpha = (\lambda, \vec q \,)$, $\lambda = 1,2$, so that 
$A_{1 \vec q} = \rho_{\vec q}$, 
$A_{2 \vec q} = j_{\vec q}$. 
Notations follow the ones of
the first paragraph of Sec.~\ref{sec:2}. 
One gets $g_{1 {\vec q} \, 1 {\vec q}} = NS_q$,
$g_{2 {\vec q} \,  2 {\vec q}} = Nv^2$,
$\Omega_{1 {\vec q} \, 2 {\vec q}} = q$,
$\Omega_{2 {\vec q} \, 1 {\vec q}} = \Omega_q^2 / q$, and all the
other elements of the matrices $g$ and $\Omega$ are zero. Because
of translational invariance, one gets 
$C_{\lambda {\vec q} \, \mu {\vec k}}(t) = 0$ unless $\vec q = \vec k$, and rotational
invariance implies that $C_{\lambda {\vec q} \, \mu {\vec q}} (t)$
depends on the modulus $q$ only. The same holds for the matrices
$\Gamma$ and $M$, so that Eq.~(\ref{eq:B6}) reduces to a two-by-two matrix
equation with $q$ appearing as a parameter. 
Since ${\cal L} \rho_{\vec q}$ 
is an element of the distinguished set of
variables, the kernels $\Gamma_{\alpha\beta}$ and $M_{\alpha\beta}
(t)$ vanish unless $\alpha = \beta = (2, {\vec q} \,)$. The latter
shall be denoted by $\Gamma_q$ and $M_q (t)$, respectively.
Equation (\ref{eq:B6}) can then be reduced to the equation of motion for the
normalized density correlator 
$\phi_q (t) = C_{1 {\vec q} \, 1 {\vec q}} (t) / NS_q$:
\bea
& &
\partial_{t}^{2} \phi_{q}(t) + \Gamma_{q} \, \partial_{t} \phi_{q}(t) 
\nonumber \\
& &
\qquad
+ \,
\Omega_{q}^{2} \, \phi_{q}(t) +
\int_{0}^{t} dt' \, M_{q}(t-t') \, \partial_{t'} \phi_{q}(t') = 0.
\label{eq:B7}
\eea
The application of the Mori-Fujisaka formalism can be summarized
as follows. The relaxation kernel $\Omega_q^2 m_q (t)$ of the
exact Eq.~(\ref{eq:GLE-v}) is approximately split into a white noise
contribution $2 \Gamma_q \delta (t)$ and a remainder $M_q (t)$,
where well defined formulas for the two contributions are
available. In this paper, the kernel $\Gamma_q$ is neglected and 
Eq.~(\ref{eq:B5}) for the kernel,
\begin{equation}
M_{q}(t) = (f_{\vec q}(t), f_{\vec q}) \, / \, N v^{2}, \quad
f_{\vec q} = f_{2 {\vec q}}, 
\label{eq:B8}
\end{equation}
shall be approximated further.

If one writes $a_{1 \vec q} = \tilde\rho_{\vec q}$ and $a_{2 \vec
q} = \tilde j_{\vec q}$, one can denote the equilibrium distribution
$w (a) = \langle g_a \rangle$ for the solvent variables as
\bea
w(a) &=& C \, \exp
\biggl\{
- (1/2N) \sum_{\vec q} [
(1 - \rho c_{q}) \, \tilde{\rho}_{\vec q} \, \tilde{\rho_{\vec q}}^{*} 
\nonumber \\
& &
\qquad \qquad \qquad \qquad \qquad
+ \,
(1/v^{2}) \, \tilde{j}_{\vec q} \, \tilde{j}_{\vec q}^{*}]
\biggr\}.
\label{eq:B9}
\eea
This is used to derive from Eqs.~(\ref{eq:B2}) and (\ref{eq:B4}):
\begin{equation}
f_{\vec q} = - i (\rho v^{2}/ N) \sum_{\vec k} ({\vec k} \cdot {\vec q}/q) \,
c_{k} \, \rho_{\vec k} \, \rho_{{\vec q}-{\vec k}} + \delta f_{\vec q},
\label{eq:B10}
\end{equation}
where $\delta f_{\vec q}$ denotes a term whose contribution to the memory
kernel turns out to be irrelevant
for the structural relaxation processes, and therefore shall be neglected.
The remaining task is the evaluation of averages
$(\rho_{\vec k}(t) \rho_{\vec p} (t), 
\rho_{\vec k^\prime} \rho_{\vec p^\prime})$, 
where the time evolution is generated by
the reduced Liouvillian. Here the original MCT ansatz is used:
$(\rho_{\vec k} (t) \rho_{\vec k^\prime}) 
(\rho_{\vec p} (t) \rho_{\vec p^\prime}) + (k \leftrightarrow p)$. 
As a result, one
gets $M_q (t) = \Omega_q^2 m_q (t)$ with the well known expression
for the mode-coupling functional in Eqs.~(\ref{eq:MCT-v}):
\bea
{\cal F}_{q}[\tilde{f}] &=& (\rho / 16 \pi^{3} q^{4})
\int d{\vec k} \, S_{q} S_{k} S_{p} 
\nonumber \\
& &
\qquad \qquad \qquad
\times \,
[{\vec q} \cdot {\vec k} \, c_{k} + {\vec q} \cdot {\vec p} \, c_{p}]^{2} \,
\tilde{f_{k}} \, \tilde{f_{p}}, 
\label{eq:BXX-1}
\eea
with $\vec p = \vec q - \vec k$.

\subsection{MCT equations for the solute molecule}

It is straightforward to generalize the preceding derivation to
systems with a diatomic solute molecule provided the latter is
considered as flexible. Therefore, let us use this modification of
the problem. It will be assumed that the kinetic energy of the
molecule is $\Sigma_a (m_a /2) \vec v_{a}^{\, 2}$ and that there is a
binding potential $V^{f \ell} (\mid \vec r_A - \vec r_B \mid)$
between the two interaction sites. The exact Eq.~(\ref{eq:GLE-u-a}) remains
valid with the frequency matrix replaced by that of the flexible
molecule $\Omega_q^{f \ell \, 2}$. It is defined via Eq.~(\ref{eq:GLE-u-b}) with
the simple velocity correlator $J_{q}^{f \ell \, a b} = \delta^{a b} (k_B
T /m_a)$.

The formulas of Sec.~\ref{subsec:B1} shall be applied with the extended set
of densities $(\rho_{\vec q}, \rho_{\vec q}^A, \rho_{\vec q}^B)$
and the corresponding longitudinal currents $(j_{\vec q}, j_{\vec
q}^A, j_{\vec q}^B)$. Thus the index consists of three bits
$\alpha = (\tau, \lambda, \vec q)$, where $\lambda = 1,2$
discriminates between densities and currents, and $\tau = O, A, B$
indicates solvent, atom $A$ and atom $B$, respectively. In
the infinite dilution limit $N \to \infty$, the equations for the
solute do not directly couple to those for the solvent. Therefore,
Eq.~(\ref{eq:B7}) remains valid and one gets a modification of 
Eq.~(\ref{eq:GLE-u-a}) for the solute:
\bea
& &
\partial_{t}^{2} {\bf F}_{q}(t) + {\bf \Gamma}_{q} \, \partial_{t} {\bf F}_{q}(t) +
{\bf \Omega}_{q}^{f \ell \, 2} \, {\bf F}_{q}(t) 
\nonumber \\
& &
\qquad \qquad
+ \,
\int_{0}^{t} dt' \, {\bf M}_{q}(t-t') \, \partial_{t'} {\bf F}_{q}(t') = {\bf 0}.
\label{eq:B12}
\eea
The relaxation kernel reads
\begin{equation}
M_{q}^{ab}(t) = (f_{\vec q}^{a}(t), f_{\vec q}^{b}) \, (m_{b} / k_{B}T), \quad
f_{\vec q}^{a} = f_{a 2 {\vec q}}.
\label{eq:B13}
\end{equation}

The determination of the Gaussian distribution of the extended
distinguished variables requires the inversion of the
three-by-three matrix $(A_{\tau 1 \vec q}, A_{\sigma 1 \vec
q})$. Making use of the infinite dilution limit $N \to \infty$,
one gets
\bea
\langle g_{a} \rangle &=& w^{O}(a) \,
\exp \biggl\{
(1/2) \sum_{\vec q} \sum_{ab} 
[(2 \rho / N) \, \delta^{ab} \, 
c_{q}^{a} \, \tilde{\rho}_{\vec q} \, \tilde{\rho}_{\vec q}^{a *} 
\nonumber \\
& &
- \,
({\bf w}_{q}^{-1})^{ab} \, \tilde{\rho}_{\vec q}^{a} \, \tilde{\rho}_{\vec q}^{b *} -
(m_{a}/k_{B}T) \, \delta^{ab} \, \tilde{j}_{\vec q}^{a} \, \tilde{j}_{\vec q}^{b *}]
\biggr\},
\label{eq:B14}
\eea
where $w^{O}(a)$ is the distribution for the solvent variables
given by Eq.~(\ref{eq:B9}). This expression is used to work
out the fluctuating force as explained in connection with 
Eq.~(\ref{eq:B10}):
\bea
f_{\vec q}^{a} &=& - i (\rho/N) (k_{B}T/m_{a}) \sum_{\vec k}
[({\vec q}-{\vec k}) \cdot {\vec q}/q] \, c_{k}^{a} \,
\rho_{\vec k}^{a} \, \rho_{{\vec q}-{\vec k}} 
\nonumber \\
& &
\qquad \qquad \qquad \qquad
+ \,
\delta f_{\vec q}^{a}.
\label{eq:B15}
\eea
As above, the contributions due to $\delta f_{\vec q}^a$ are
neglected and the remaining pair correlations are factorized. This
leads to
\begin{equation}
{\bf M}_{q}(t) = {\bf \Omega}_{q}^{f \ell \, 2} \, {\bf m}_{q}(t),
\label{eq:B16}
\end{equation}
where the functional for the kernel in Eq.~(\ref{eq:MCT-u-a}) reads
\bea
{\cal F}^{ab}_{q}[\tilde{\textit{\textbf f}}, \tilde{f}] &=& q^{-2} 
\sum_{c} w^{ac}_{q} \, (\rho / 8 \pi^{3})
\int d{\vec k} \, ({\vec q} \cdot {\vec p} / q)^{2} 
\nonumber \\
& &
\qquad \qquad \qquad \qquad
\times \,
S_{p} \, c_{p}^{c} \, c_{p}^{b} \, \tilde{f}_{k}^{cb} \, \tilde{f}_{p},
\label{eq:BXX-2}
\eea
with $\vec p$ abbreviating $\vec q - \vec k$. Neglecting the
friction term ${\bf \Gamma}_{q}$ in Eq.~(\ref{eq:B12}), 
the MCT equations (\ref{eq:GLE-u}) and (\ref{eq:MCT-u}) 
are derived for the flexible molecule.

Obviously, a theory for a rigid molecule can be obtained from one
for a flexible molecule only within a quantum-mechanical
approach. One has to consider the case where excitation energies
for translational and rotational motion are small compared to the
thermal energy, while the energies for
vibrational excitations are large. Let
us assume, that the formulas can be obtained by replacing all
equilibrium averages in the preceding equations by the correct
quantum mechanical ones, where the latter can be evaluated for the
classical molecule model with five degrees of freedom. This
amounts to replacing structure functions by the classical
quantities, in particular the replacement of ${\bf \Omega}_q^{f
\ell}$ by ${\bf \Omega}_q$. Thereby Eq.~(\ref{eq:B12}) produces 
Eqs.~(\ref{eq:GLE-u}).

\begin{figure}
\centerline{\scalebox{0.55}{\includegraphics{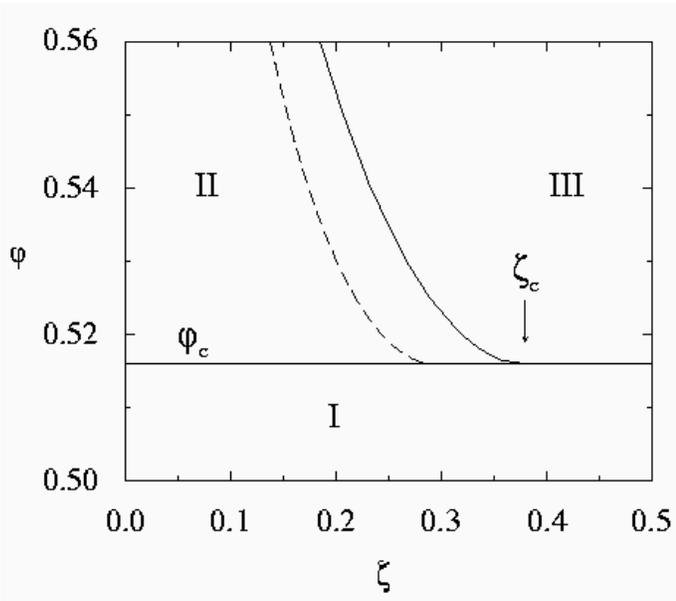}}}
\caption{Phase diagram of a dilute solute of symmetric dumbbell
molecules with elongation $\zeta$ consisting of two fused hard spheres
which are immersed in a hard-sphere system (HSS) with packing
fraction $\varphi$. The horizontal line marks the liquid-glass
transition at the critical packing fraction
$\varphi_c = 0.516$. The other full line is the curve $\zeta_c
(\varphi)$ of critical elongations for a type-$A$ transition
between phases II and III. In phase II dipole fluctuations of the
solute relax to zero for long times, while they are frozen in
phase III. The value $\zeta_c = \zeta_c (\varphi_c) = 0.380$ 
is marked by an arrow. The dashed
line is the corresponding transition curve calculated 
in Ref.~\protect\cite{Franosch98b}; 
it terminates at $\zeta_c^t = 0.297$.}
\label{fig:phase}
\end{figure}

\newpage\noindent

\begin{figure}
\centerline{\scalebox{0.85}{\includegraphics{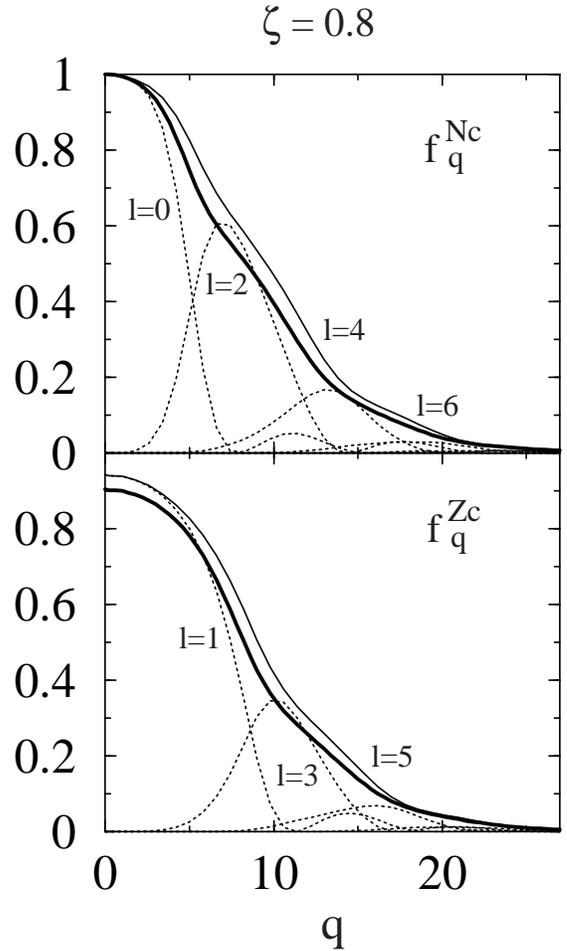}}}
\caption{
Nonergodicity parameters $f_q^{x \,c}$ (heavy full lines)
for the molecule's arrested number-density
fluctuations $(x = N)$ and 
``charge''-density fluctuations $(x = Z)$ 
for the critical packing fraction
$\varphi = \varphi_c$.
The elongation parameter $\zeta = 0.80$ is
representative for strong steric hindrance for reorientational
motion. The light full lines are evaluated with 
Eq.~(\ref{eq:Fab-site-tensor-2}) with the
nonergodicity parameters $f^c (q \ell 0)$ obtained in 
Ref.~\protect\cite{Franosch97c} from a theory based on a
tensor-density description. The dotted lines show the
contributions to Eq.~(\ref{eq:Fab-site-tensor-2}) 
from different angular-momentum index
$\ell$. Here and in the following figures the diameter of the
spheres is used as unit of length, $d = 1$. }
\label{fig:fq-strong}
\end{figure}

\newpage\noindent

\begin{figure}
\centerline{\scalebox{0.85}{\includegraphics{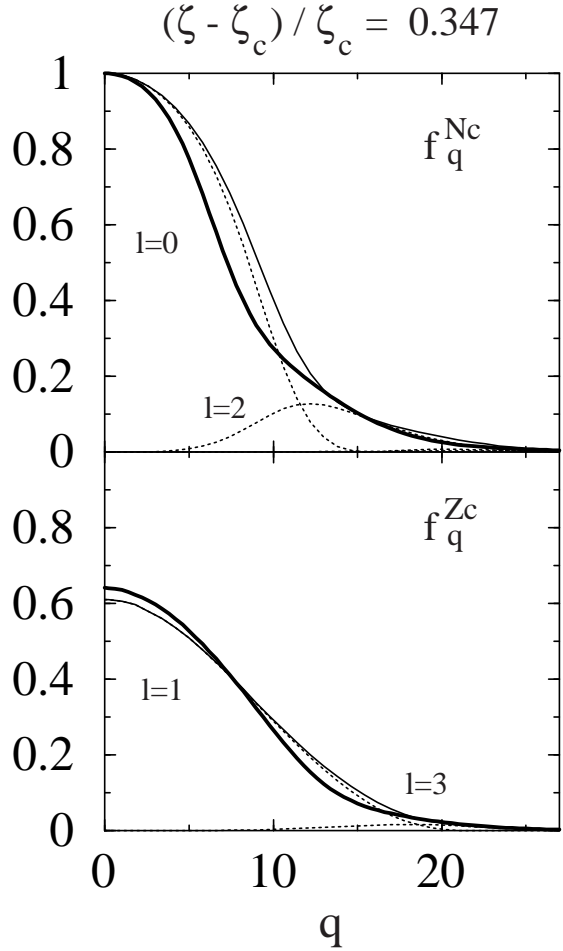}}}
\caption{Results as in Fig.~\protect\ref{fig:fq-strong} 
but for small elongations which are
representative for weak steric hindrance for reorientational
motion. The relative distance from the transition point between
phases II and III is $(\zeta - \zeta_c) / \zeta_c = 0.347$. The
heavy full line shows the result of the present theory for 
$\zeta = 0.512$. The light full line
shows the result for $\zeta=0.400$ based on 
Ref.~\protect\cite{Franosch97c}.}
\label{fig:fq-weak}
\end{figure}

\newpage\noindent

\begin{figure}
\centerline{\scalebox{0.85}{\includegraphics{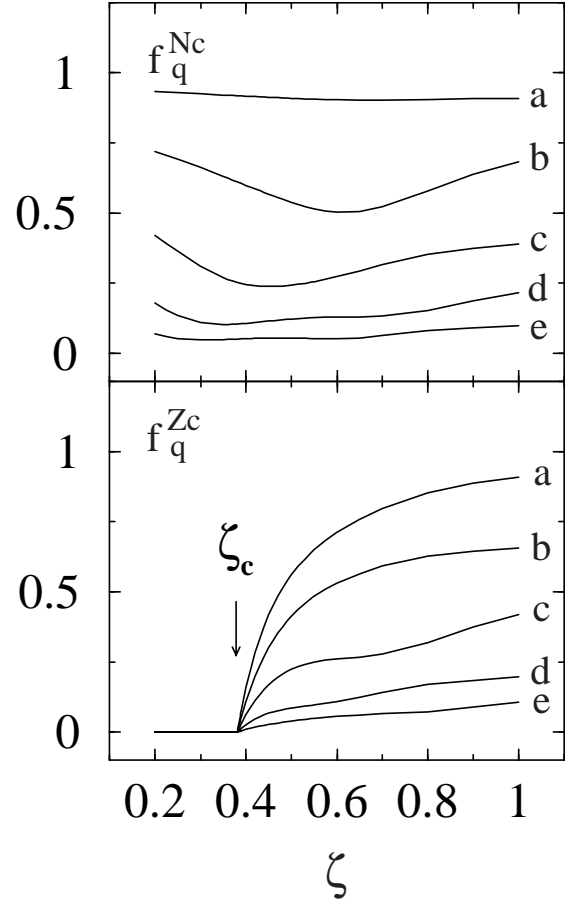}}}
\caption{Critical nonergodicity parameters $f_q^{x \, c}$ for the solute
as function of the elongation parameter $\zeta$ for the 
wave numbers $q = 3.4 (a), \, 7.0 (b), \, 10.6
(c), \, 14.2 (d)$, and 17.4 $(e)$. The arrow 
marks the transition point from phase II to phase III at $\zeta_c
= 0.380$. }
\label{fig:fq-vs-zeta}
\end{figure}

\newpage\noindent

\begin{figure}
\centerline{\scalebox{0.65}{\includegraphics{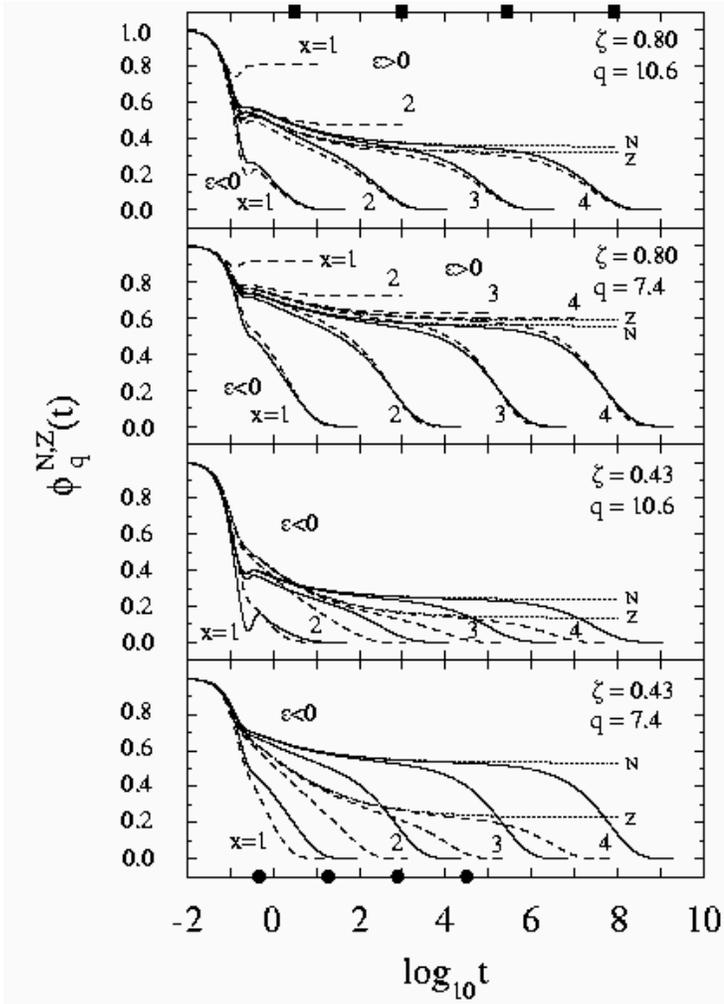}}}
\caption{Correlators $\phi_q^N (t)$ (solid lines)
and $\phi_q^Z (t)$ (dashed lines)
for two intermediate wave numbers $q$ as function of the
logarithm of time $t$. The decay curves at the
critical packing fraction $\varphi_c$ for number-density and
``charge''-density correlators are shown as dotted lines and marked
by $N$ and $Z$, respectively. Only a few solutions of glass states
are shown for $\phi_q^Z (t)$ in order to avoid overcrowding of the
figure. The distance parameter is 
$\epsilon = (\varphi - \varphi_c) / \varphi_c = \pm 10^{-x}$.
The full circles and squares mark the characteristic
times $t_{\sigma}$ and $t'_{\sigma}$, respectively, according to
Eq.~(\protect\ref{eq:t-sigma}) 
for $x = 1$, 2, 3 and 4.
The unit of time is chosen here and in the
following figures such that the thermal velocity of the solvent
reads $v = 1$.} 
\label{fig:NN-ZZ-t}
\end{figure}

\newpage\noindent

\begin{figure}
\centerline{\scalebox{0.80}{\includegraphics{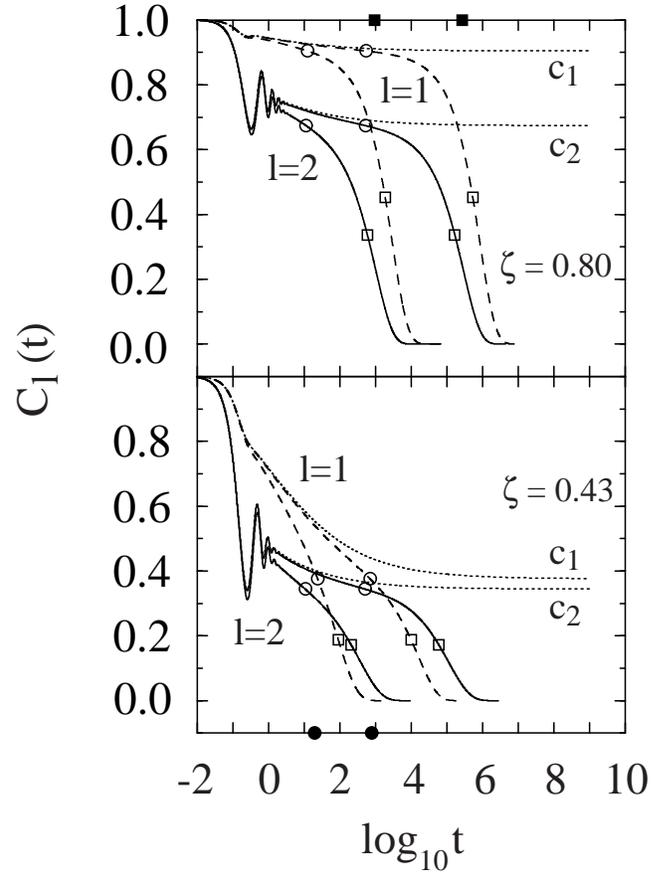}}}
\caption{Reorientational correlators $C_\ell(t)$ for $\ell = 1$
(dashed lines) and $\ell = 2$ (full lines) 
for two elongations $\zeta$. The correlators for the
critical packing fraction $\varphi = \varphi_c$ are shown as
dotted lines marked with $c_\ell$. The distance parameters
$\epsilon = (\varphi - \varphi_c) / \varphi_c$ are -0.01 (faster
decay) and -0.001 (slower decay). 
The full circles and squares mark the
corresponding time scales $t_\sigma$ and $t_\sigma^\prime$,
respectively, from Eq.~(\protect\ref{eq:t-sigma}). 
The open circles and squares on the curves mark the characteristic
time scales $\tau_{\beta}^{\ell}$ and $\tau_{\alpha}^{\ell}$,
respectively, 
defined by $C_{\ell}(\tau_{\beta}^{\ell}) = f_{\ell}^{c}$ and
$C_\ell (\tau_\alpha^\ell) = f_\ell^c / 2$. }
\label{fig:C1-C2}
\end{figure}

\newpage\noindent

\begin{figure}
\centerline{\scalebox{0.80}{\includegraphics{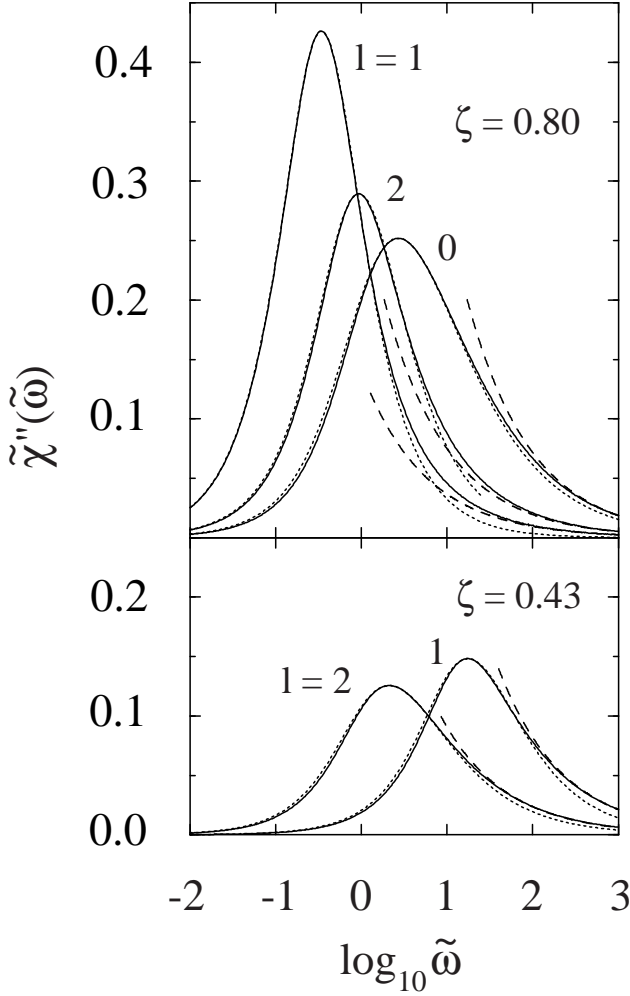}}}
\caption{Susceptibility master spectra 
$\tilde{\chi}^{\prime \prime} (\tilde{\omega})$ of
the $\alpha$-process as function of the logarithm of the rescaled
frequency $\tilde \omega = \omega t_\sigma^\prime$ (see text). 
Upper panel: 
the curves $\ell = 1$ and 2 refer to the response for the dipole
and quadrupole, respectively, for elongation $\zeta = 0.80$. The
curve $\ell = 0$ refers to the susceptibility master spectrum of 
the dimensionless longitudinal elastic modulus $m_{q=0} (t)$ of the HSS.
The dashed lines exhibit the von Schweidler tails, Eq.~(\ref{eq:von-b}). 
The dotted lines are fits by Kohlrausch spectra 
$\tilde \chi_K^{\prime \prime}(\tilde{\omega})$ 
with stretching exponents $\beta = 0.97$, 0.88 and 0.63
chosen for $\ell = 1, 2$ and 0, respectively, so that 
the maximum and the full width
at the half maximum $W$ in decades of 
$\tilde \chi_K^{\prime \prime}(\tilde{\omega})$ agree
with those of $\tilde \chi^{\prime \prime}(\tilde{\omega})$. 
The position of the susceptibility maximum is $\tilde \omega_{\max} = 0.337$ (0.927,
2.69) and the width is $W = 1.17$ (1.28, 1.76) for $\ell = 1$ (2, 0).
Lower panel: corresponding results for $\zeta = 0.43$.
The stretching exponent $\beta$, the maximum position
$\tilde \omega_{\max}$ and the width $W$ for $\ell=1$ ($\ell=2$) are
$\beta = 0.79$ (0.71), 
$\tilde \omega_{\max} = 17.5$ (2.11)
and
$W = 1.42$ (1.57), respectively.}
\label{fig:alpha-spectra}
\end{figure}

\newpage\noindent

\begin{figure}
\centerline{\scalebox{0.70}{\includegraphics{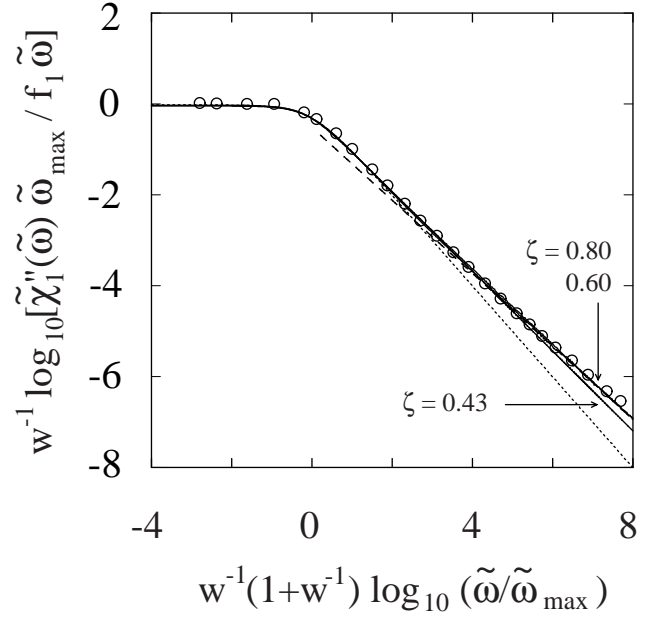}}}
\caption{Double logarithmic presentation of the normalized
fluctuation spectra of
the dipole-reorientation $\alpha$-processes, 
$\tilde{\chi}^{\prime \prime}_{1} (\tilde{\omega})
\tilde{\omega}_{\max} / f_{1}^{c} \tilde{\omega}$, as function of 
$\tilde{\omega} / \tilde{\omega}_{\max}$.
Here $\tilde{\omega}_{\max}$ denotes the position of the susceptibility
maximum. Following Dixon {\it et al.} \protect\cite{Dixon90} the vertical axis
is rescaled by $w^{-1}$ and the horizontal one by $w^{-1} (1 +
w^{-1})$, where $w = W / W_D$ is the ratio of the logarithmic full width
at half maximum $W$ of the susceptibility peak to the same quantity $W_D$ of
a Debye-peak. The open circles reproduce some of the dielectric-loss
results for glycerol \protect\cite{Dixon90}. 
The three full lines from the top
are the results for $\zeta = 0.80$, 0.60 and 0.43,
successively,
although the upper two curves cannot be distinguished within the 
resolution of the figure.
The dotted and the dashed lines exhibit 
the Kohlrausch fit with the stretching exponent
$\beta = 0.97$ and the von-Schweidler-law tail, respectively,
for the $\zeta = 0.80$--spectrum.}
\label{fig:Nagel-plot}
\end{figure}

\newpage\noindent

\begin{figure}
\centerline{\scalebox{0.80}{\includegraphics{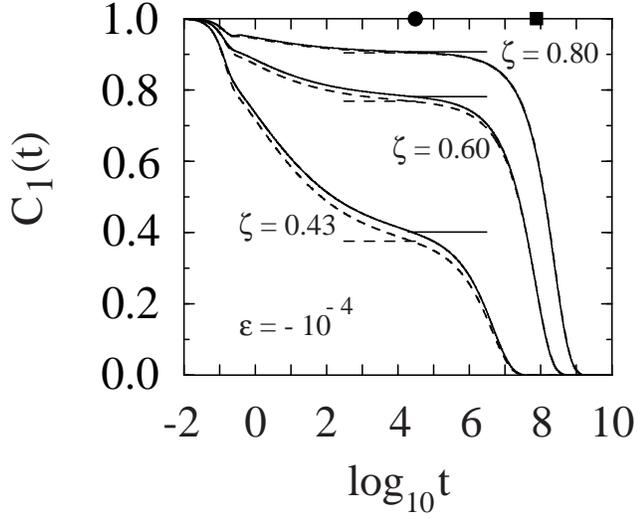}}}
\caption{Dipole correlators $C_1$(t) for the distance parameter
$\epsilon = - 10^{-4}$ for three elongations $\zeta$. The full circle
and square indicate the times $t_\sigma$ and $t_\sigma^\prime$,
respectively, from Eq.~(\ref{eq:t-sigma}).
The dashed lines are calculated from 
Eqs.~(\ref{eq:GLE-C1}) and (\ref{eq:MCT-MSD-C1}). 
The plateaus $f_1^c = 0.905$ (0.769, 0.376) for
$\zeta = 0.80$ (0.60, 0.43) are shown by dashed horizontal lines. The
full lines exhibit $C_1(t)$ for the same states, but evaluated from
equations derived in analogy to 
Eqs.~(\ref{eq:GLE-C2})--(\ref{eq:m2-final}),
and their plateaus 
$f_1^c = 0.907$ (0.782, 0.402) are indicated by full horizontal lines. }
\label{fig:compare-C1}
\end{figure}

\newpage\noindent

\begin{figure}
\centerline{\scalebox{0.55}{\includegraphics{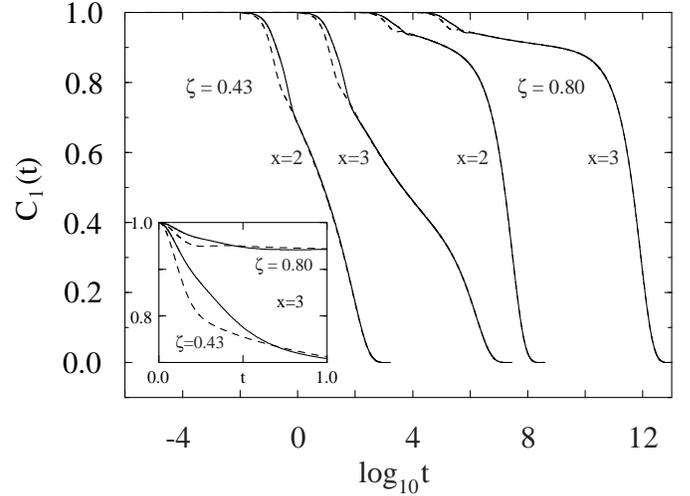}}}
\caption{Dipole correlators $C_1 (t)$ for a dumbbell of two fused
hard spheres of diameters $d$ and distance $\zeta d$ between the
centers moving in a liquid of hard spheres with diameter $d$
for a distance parameter
$(\varphi-\varphi_{c})/\varphi_{c} = - 10^{-x}$.
The dashed lines reproduce the results from Fig.~\ref{fig:C1-C2} and 
refer to a symmetric molecule with masses of the two atoms being 
equal to the mass $m$ of the solvent particles $m_A = m_B = m$. 
The full lines
exhibit the results for an asymmetric dumbbell with 
$m_A = 10 m$, $m_B = m$. 
In the main frame the results for different states
are successively shifted horizontally by 2 decades in order
to avoid overcrowding.
The inset shows the transient dynamics 
on a linear time axis. }
\label{fig:C1-inertia}
\end{figure}

\end{document}